\documentclass[conference,compsoc]{./sty/IEEEtran}
\usepackage[square,comma,numbers,sort&compress]{natbib}
\usepackage{silence}
\usepackage[hyphens]{url}
\usepackage[breaklinks,colorlinks]{hyperref}
\usepackage[usenames,dvipsnames]{xcolor}
\hypersetup{citecolor=blue,linkcolor=blue}
\usepackage{amsmath,amsopn,amssymb}
\usepackage{endnotes,microtype,xspace,graphicx,fancyvrb,multirow}
\usepackage{booktabs}
\usepackage{array,underscore,relsize}
\usepackage[T1]{fontenc}
\usepackage{times}
\usepackage{fancyhdr}
\usepackage{enumitem}
\usepackage[labelfont=bf,font=small,skip=5pt]{caption}
\usepackage{makecell}
\usepackage{subcaption}
\usepackage{comment}
\usepackage{tabularx}
\usepackage{threeparttable}

\usepackage[ruled,vlined]{algorithm2e}

\SetCommentSty{mycommfont}
\usepackage{fdsymbol}

\ifdefined\DIFF
\fancypagestyle{firstpage}{%

}
\else
\fi

\usepackage{mdframed}
\ifdefined\DIFF
\mdfdefinestyle{AddFrame}{%
  nobreak=true, %
  linecolor=ao,
  linewidth=2pt,
  innertopmargin=0pt,
  innerbottommargin=0pt,
  innerrightmargin=0pt,
  innerleftmargin=0pt,
}
\else
\mdfdefinestyle{AddFrame}{%
  nobreak=true, %
  linewidth=0pt,
  innertopmargin=0pt,
  innerbottommargin=0pt,
  innerrightmargin=0pt,
  innerleftmargin=0pt,
}
\fi

\usepackage{amssymb}%
\usepackage{pifont}%
\usepackage{fp}
\usepackage{siunitx}

\usepackage{balance}

\newcommand{\sys}{\mbox{\textsc{TikTag}}\xspace}

\newcommand{\objvuln}{$\cc{obj}_{\text{vuln}}$\xspace}
\newcommand{\objtarget}{$\cc{obj}_{\text{target}}$\xspace}

\newcommand{\BL}[1]{}
\newcommand{\JH}[1]{}
\newcommand{\JB}[1]{}

\newcommand{\cc}[1]{\mbox{\smaller[0.5]\texttt{#1}}}

\fvset{fontsize=\scriptsize,xleftmargin=8pt,numbers=left,numbersep=5pt}

\input{code/fmt}

\def\Snospace~{\S{}}

\newif\ifdraft\drafttrue
\newif\ifnotes\notestrue
\ifdraft\else\notesfalse\fi

\input{glyphtounicode}
\newcolumntype{R}[1]{>{\raggedleft\let\newline\\\arraybackslash\hspace{0pt}}p{#1}}

\newcommand{\squishlist}{
\begin{itemize}[noitemsep,nolistsep]
  \setlength{\itemsep}{-0pt}
}
\newcommand{\squishend}{
  \end{itemize}
}

\usepackage{tikz}
\newcommand*\C[1]{%
\begin{tikzpicture}[baseline=(C.base)]
\node[draw,circle,inner sep=0.2pt](C) {#1};
\end{tikzpicture}}

\usepackage{xstring}
\newcommand{\PP}[1]{
\vspace{2px}
\noindent{\bf \IfEndWith{#1}{.}{#1}{#1.}}
}

\gdef\therev{105431a}
\gdef\thedate{2024-06-11 17:00:19 +0900}

\graphicspath{{./images/}}

\begin{document}

\title{\sys: Breaking ARM's Memory Tagging Extension with Speculative Execution}

\ifdefined\DRAFT
 \pagestyle{fancyplain}
 \lhead{Rev.~\therev}
 \rhead{\thedate}
 \cfoot{\thepage\ of \pageref{LastPage}}
\fi

\author{\IEEEauthorblockN{Juhee Kim}
\IEEEauthorblockA{Seoul National University\\
kimjuhi96@snu.ac.kr}
\and
\IEEEauthorblockN{Jinbum Park}
\IEEEauthorblockA{Samsung Research\\
jinb.park@samsung.com}
\and
\IEEEauthorblockN{Sihyeon Roh}
\IEEEauthorblockA{Seoul National University\\
sihyeonroh@snu.ac.kr}
\and
\IEEEauthorblockN{Jaeyoung Chung}
\IEEEauthorblockA{Seoul National University\\
jjy600901@snu.ac.kr}
\and
\IEEEauthorblockN{Youngjoo Lee}
\IEEEauthorblockA{Seoul National University\\
youngjoo.lee@snu.ac.kr}
\and
\IEEEauthorblockN{Taesoo Kim}
\IEEEauthorblockA{Samsung Research and \\ Georgia Institute of Technology\\
taesoo@gatech.edu}
\and
\IEEEauthorblockN{Byoungyoung Lee}
\IEEEauthorblockA{Seoul National University\\
byoungyoung@snu.ac.kr}
}

\date{}
\maketitle

\begin{abstract}

ARM Memory Tagging Extension (MTE) is a new hardware feature
introduced in ARMv8.5-A architecture, aiming to detect memory
corruption vulnerabilities.
The low overhead of MTE makes it an attractive solution to mitigate
memory corruption attacks in modern software systems and is
considered the most promising path forward for improving C/C++
software security.
This paper explores the potential security risks posed by speculative
execution attacks against MTE.
Specifically, this paper identifies new \sys gadgets capable of
leaking the MTE tags from arbitrary memory addresses through
speculative execution.
With \sys gadgets, attackers can bypass the probabilistic defense of
MTE, increasing the attack success rate by close to 100\%.
We demonstrate that \sys gadgets can be used to bypass MTE-based
mitigations in real-world systems, Google Chrome and the Linux
kernel.
Experimental results show that \sys gadgets can successfully leak an
MTE tag with a success rate higher than 95\% in less than 4 seconds.
We further propose new defense mechanisms to mitigate the security
risks posed by \sys gadgets.

\end{abstract}

\section{Introduction}
\label{s:intro}

Memory corruption vulnerabilities present significant security
threats to computing systems.
Exploiting a memory corruption vulnerability, an attacker corrupts
the data stored in a memory, hijacking the control flow or crafting
the data of the victim.
Such exploitation allows the attacker to execute arbitrary code,
escalate its privilege, or leak security-sensitive data, critically
harming the security of the computing system.

In response to these threats, ARM Memory Tagging Extension (MTE) has
recently been proposed since ARMv8.5-A architecture, which is a new
hardware extension to mitigate memory corruption attacks.
Technically, MTE provides two hardware primitive operations,
(i)~\emph{tag} and (ii)~\emph{tag check}.
A tag operation assigns a tag to a memory location (i.e., a 4-bit tag
to each 16-byte memory).
Then a tag check operation is performed when accessing the memory,
which compares two tags, one embedded within the pointer to access the
memory and the other associated with the memory location to-be-accessed.
If these two tags are the same, the access is allowed.
Otherwise, the CPU raises a fault.

Using MTE, various mitigation techniques can be developed depending
on which tag is assigned or which memory regions are tagged.
For instance, MTE-supported memory allocators, such as Android
Scudo~\cite{scudo} and Chrome PartitionAlloc~\cite{partition}, assign
a random tag for all dynamically allocated memory.
Specifically, a memory allocator is modified to assign a random tag
for each allocation.
Then, a pointer to this allocated memory embeds the tag, and as the
pointer is propagated, the tag is accordingly propagated together.
When any dynamically allocated memory is accessed, a tag check
operation is enforced.
As the tags are randomly assigned at runtime, it is difficult for the
attacker to correctly guess the tag. Thus tag check operation would
statistically detect memory corruptions.

MTE introduces significant challenges for attackers to exploit the
memory corruption vulnerability.
This is because MTE-based solutions detect a violation behavior
close to the root cause of spatial and temporal memory corruptions.
Specifically, since MTE ensures that the relationship between a
pointer and a memory location is not corrupted, it promptly detects
the corruptions---i.e., MTE promptly detects the moment when
out-of-bounds access takes place in a heap-overflow vulnerability or
when a dangling pointer is dereferenced in use-after-free.
This offers strong security advantages to MTE, particularly compared
to popular mitigation techniques such as CFI~\cite{cfi,
edge-cfi,android-cfi}, which does not detect memory corruption but
detects control-flow hijack behavior (i.e., an exploitation
behavior).
For these reasons, MTE is considered \emph{the most promising path
forward for improving C/C++ software security} by many security
experts~\cite{mte-blog,arm-blog}, since its first adoption with the
Pixel~8 device in October 2023.

In this paper, we study if MTE provides the security assurance as it
is promised.
In particular, we analyzed if speculative execution attacks can be a
security threat to breaking MTE.
To summarize our results, we found that speculative execution attacks
are indeed possible against MTE, which severely harms the security
assurance of MTE.
We discovered two new gadgets, named \sys-v1 and \sys-v2, which can
leak the MTE tag of an arbitrary memory address.
Specifically, \sys-v1 exploits the speculation shrinkage of the branch
prediction and data prefetchers, and \sys-v2 exploits the
store-to-load forwarding behavior.

To demonstrate the exploitability of real-world MTE-based mitigations,
we developed two real-world attacks having distinctive attack
surfaces: Google Chrome and the Linux kernel.
Our evaluation results show that \sys gadgets can leak MTE tags with a
success rate higher than 95\% in less than 4 seconds.
We further propose mitigation schemes to prevent the
exploitation of \sys gadgets while retaining the benefits of using MTE.

Compared to the previous works on MTE side-channels~\cite{pz-mte,
  sticky-tags}, we think this paper makes unique contributions for the
following reasons.
First, Project Zero at Google reported that they were not able to find
speculative tag leakage from the MTE mechanisms~\cite{pz-mte}.
They concluded that speculative MTE check results do not induce
distinguishable cache state differences between the tag check success
and failure.
In contrast, we found that tag checks indeed generate the cache state
difference in speculative execution.

Another independent work, \cc{StickyTags}~\cite{sticky-tags},
discovered an MTE tag leakage gadget, which is one example of the
\sys-v1 gadget, and suspected that the root cause is in the memory
contention on spurious tag check faults.
On the contrary, this paper performed an in-depth analysis, which
identified that the speculation shrinkage in branch prediction and
data prefetchers are the root cause of the \sys-v1 gadget.
This paper additionally reports new MTE tag leakage gadgets,
specifically the variants of \sys-v1 gadget and the new \sys-v2
gadget, along with developing exploitation against Chrome and the
Linux kernel.
Furthermore, this paper proposes new defense mechanisms to prevent
\sys gadgets from leaking MTE tags, both at hardware and software
levels.

At the time of writing, MTE is still in the early stages of adoption.
Considering its strong security advantage, it is expected that a large
number of MTE-based mitigations (e.g., sensitive data
protection~\cite{gv-protection, knox-rkp} and data-flow
integrity~\cite{dfi, kenali, hakc}) is expected to be deployed in the
near future on MTE-supporting devices (e.g., Android mobile phones).
As such, the results of this paper, particularly in how \sys gadgets
are constructed and how MTE tags can be leaked, shed light on how
MTE-based solutions should be designed or how CPU should be
implemented at a micro-architectural level.
We have open-sourced \sys gadgets
at~\cc{https://github.com/compsec-snu/tiktag} to help the community
understand the MTE side-channel issues.

\PP{Responsible Disclosure}
We reported MTE tag leakage gadgets to ARM in November 2023. 
ARM acknowledged and publicly disclosed the issue in December
2023~\cite{arm-tag-leak}.
Another research group reported a similar issue to ARM and published
their findings~\cite{sticky-tags}, which were conducted independently
from our work.
We reported the speculative vulnerabilities in Google Chrome V8 to the
Chrome Security Team in December 2023.
They acknowledged the issues but decided not to fix the
vulnerabilities because the V8 sandbox is not intended to guarantee
the confidentiality of memory data and MTE tags.
Since the Chrome browser currently does not enable its MTE-based
defense by default, we agree with their decision to some extent.
However, we think that browser security can be improved if MTE-based
defenses are deployed with the countermeasures we
suggest~(\autoref{s:chrome-mitigation}).
We also reported the MTE oracles in the Pixel 8 device to the Android
Security Team in April 2024.
Android Security Team acknowledged the issue as a hardware flaw of
Pixel 8, decided to address the issue in Android's MTE-based defense,
and awarded a bounty reward for the report.

\section{Background}
\label{s:background}

\subsection{Memory Tagging Extension}
Memory Tagging Extension (MTE)~\cite{mte} is a hardware extension to
prevent memory corruption attacks, available since ARMv8.5-A
architecture.
MTE has been recently adopted by Pixel 8~\cite{pixel} since October
2023.
MTE assigns a 4-bit tag for each 16 bytes of memory and stores the
tag in the unused upper bits of a pointer.
During memory access, the tag in the pointer is checked against the
tag assigned for the memory region.
If the tags match, access is permitted; otherwise, the CPU raises a
tag check fault (TCF).

MTE offers three modes---\emph{synchronous}, \emph{asynchronous}, and
\emph{asymmetric}---to balance performance and security.
Synchronous mode provides the strongest security guarantee, where the
tag check fault is synchronously raised at the faulting load/store
instruction.
Asynchronous mode offers the best performance, where the tag check
fault is asynchronously raised at context switches.
Asymmetric mode strikes a balance between performance and security, with
load instructions operating in synchronous mode and store
instructions in asynchronous mode.

Based on MTE, various mitigation schemes can be developed.
\emph{deterministic tagging} assigns a globally known tag to
each allocation.
This approach can deterministically isolate memory
regions~\cite{color} or detect bounded spatial memory
corruptions~\cite{sticky-tags}. 
\emph{random tagging}, on the other hand, assigns a random tag
generated at allocation time.
This approach probabilistically prevents spatial and temporal memory
errors at per-allocation granularity, with a maximum detection rate
of 15/16 (i.e., 1/16 chance of tag collision).
Unlike deterministic tagging, random tagging does not reveal the tag
information to attackers, requiring them to guess the tag to exploit
memory corruption vulnerabilities.
Consequently, random tagging is commonly adopted in real-world
allocators (e.g., Android Scudo allocator~\cite{scudo}, Chrome
PartitionAlloc~\cite{partition}) and Linux Hardware Tag-Based
KASAN~\cite{hw-kasan}.

\subsection{Speculative Execution Attack}
A speculative execution attack is a class of attacks that exploit the
CPU's speculative behaviors to leak sensitive
information~\cite{spectre,meltdown,foreshadow,lvi,spoiler,retbleed,fallout,ridl}.
Spectre~\cite{spectre} and Meltdown~\cite{meltdown} are well-known
speculative execution attacks, where the attacker speculatively
executes the victim code to load data that cannot be accessed during
the normal execution (e.g., out-of-bounds access).
If the speculatively loaded data affects the cache, the attacker can
infer its value by observing the cache state (e.g., cache hit/miss
based on access latency).
Such speculative information leakage typically requires two attacker's
capabilities: i) controlling the cache state by flushing or evicting
cache sets before the victim accesses the data, and ii) measuring
the time precisely enough to discern cache hits and misses.
Recent studies have extended speculative execution attacks to bypass
hardware security features such as Address Space Layout Randomization
(ASLR)~\cite{aslr} and Pointer Authentication Code (PAC)~\cite{pac}.

\section{Threat Model}
\label{s:threat}

We consider a threat model where the target system employs Memory
Tagging Extension (MTE)~\cite{mte} to prevent memory corruption.
The allocator in the target system tags each allocation with a
\emph{random tag}, and the tag is checked on every memory access.
We assume random tagging since it is architecturally designed to
improve security~\cite{mte} and commonly developed in real-world
MTE-enabled systems (e.g., Android scudo allocator~\cite{scudo},
Chrome PartitionAlloc~\cite{partition}, and Linux Hardware Tag-Based
KASAN~\cite{hw-kasan}).

We assume that the attacker possesses knowledge of the memory 
corruption vulnerabilities in the target system, and aims to exploit
the vulnerabilities to gain unauthorized access to the system.
From the attacker's perspective, triggering the vulnerabilities
imposes a high probability of crashing the target process with a tag
check fault, which notifies the system administrator of the attack.
We further detail the specific threat model in real-world attack
scenarios~(\autoref{s:attack}).

\section{Finding Tag Leakage Gadgets}
\label{s:discover}
The security of MTE random tag assignment relies on the
confidentiality of the tag information per memory address.
If the attacker can learn the tag of a specific memory address, it can
be used to bypass MTE---e.g., exploiting memory corruption only when
the tag match is expected.
In this section, we present our approach to discovering MTE tag
leakage gadgets. 
We first introduce a template for an MTE tag leakage
gadget~(\autoref{s:discover:template}) and then present a
template-based fuzzing to discover MTE tag leakage gadgets
(\autoref{s:discover:fuzzing}).

\begin{figure}[t]
  \centering
  \includegraphics[width=0.8\columnwidth]{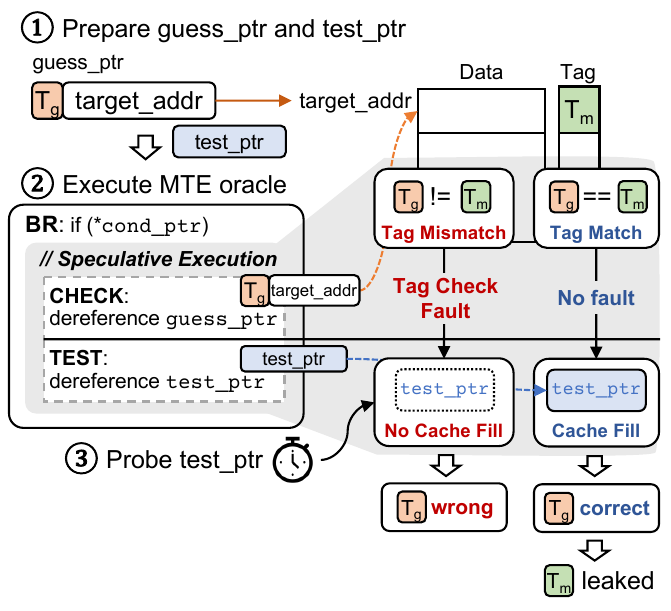}
  \caption{An MTE tag leakage template}
  \label{fig:template}
\end{figure}

\subsection{Tag Leakage Template}
\label{s:discover:template}
We first designed a template for a speculative MTE tag leakage gadget,
which allows the attacker to leak the tag of a given memory address
through speculative execution~(\autoref{fig:template}).
The motivation behind the template is to trigger MTE tag checks
in a speculative context and observe the cache state after the
speculative execution.
If there is any difference between tag match and mismatch, the
attackers can potentially leak the tag check results and infer the tag
value.
Since tag mismatch during speculative execution is not raised as an
exception, such an attempt is not detected.

We assume the attacker aims to leak the tag \cc{Tm} assigned to
\cc{target_addr}.
To achieve this, the attacker prepares two pointers: \cc{guess_ptr}
and \cc{test_ptr}~(\C{1}).
\cc{guess_ptr} points to \cc{target_addr} while embedding a tag
\cc{Tg}---i.e., \cc{guess_ptr = (Tg<<56)|(target_addr \&
  \textasciitilde(0xff<<56))}.
\cc{test_ptr} points to an attacker-accessible, uncached address
with a valid tag.

Next, the attacker executes the template with \cc{guess_ptr} and
\cc{test_ptr}~(\C{2}).
The template consists of three components in order: \cc{BR},
\cc{CHECK}, and \cc{TEST}.
\cc{BR} encloses \cc{CHECK} and \cc{TEST} using a conditional
branch, ensuring that \cc{CHECK} and \cc{TEST} are speculatively
executed.
In \cc{CHECK}, the template executes a sequence of memory
instructions to trigger MTE checks.
In \cc{TEST}, the template executes an instruction updating the cache
status of \cc{test_ptr}, observable by the attacker later.

Our hypothetical expectation from this template is as follows:
The attacker first trains the branch predictor by executing the
template with \cc{cond_ptr} storing 1 and \cc{guess_ptr} containing a
valid address and tag.
After training, the attacker executes the template with \cc{cond_ptr}
storing 0 and \cc{guess_ptr} pointing to \cc{target_addr} with a
guessed tag, causing speculative execution of \cc{CHECK} and
\cc{TEST}.
If the MTE tag matches in \cc{CHECK}, the CPU would continue to
speculatively execute \cc{TEST}, accessing \cc{test_ptr} and filling
its cache line.
If the tags do not match, the CPU may halt the speculative execution
of \cc{TEST}, leaving the cache line of \cc{test_ptr} unfilled.
Consequently, the cache line of \cc{test_ptr} would not be filled.
After executing the template, the attacker can measure the access
latency of \cc{test_ptr} after execution, and distinguish the cache
hit and miss, leaking the tag check result~(\C{3}).
The attacker can then brute-force the template executions with all
possible \cc{Tg} values to eventually leak the correct tag of
\cc{target_addr}.

\PP{Results}
We tested the template on real-world ARMv8.5 devices, Google Pixel 8
and Pixel 8 pro.
We varied the number and type of memory instructions in \cc{CHECK} and
\cc{TEST}, and observed the cache state of \cc{test_ptr} after
executing the template.
As a result, we identified two speculative MTE leakage gadgets,
\sys-v1~(\autoref{s:gadget1}) and \sys-v2~(\autoref{s:gadget2}) that
leak the MTE tag of a given memory address in both Pixel 8 and Pixel 8
pro.

\subsection{Tag Leakage Fuzzing}
\label{s:discover:fuzzing}

To automatically discover MTE tag leakage gadgets, we developed a
fuzzer in a similar manner to the Spectre-v1 fuzzers~\cite{revizor}.
The fuzzer generates test cases consisting of a sequence of assembly
instructions for the speculatively executed blocks in the tag leakage
template (i.e., \cc{CHECK} and \cc{TEST}).
The fuzzer consists of the following steps:

Based on the template, the fuzzer first allocates memory for
\cc{cond_ptr}, \cc{guess_ptr}, and \cc{test_ptr}.
\cc{cond_ptr} and \cc{guess_ptr} point to a fixed 128-byte memory
region individually.
\cc{test_ptr} points to a variable 128-byte aligned address from a
4KB memory region initialized with random values.
Then, the fuzzer randomly picks two registers to assign \cc{cond_ptr}
and \cc{guess_ptr} from the available registers (i.e., \cc{x0-x28}).
The remaining registers hold a 128-byte aligned address within a 4KB
memory region or a random value.

The fuzzer populates \cc{CHECK} and \cc{TEST} blocks using a
predefined set of instructions (i.e., \cc{ldr}, \cc{str}, \cc{eor},
\cc{orr}, \cc{nop}, \cc{isb}) to reduce the search space.
Given an initial test case, the fuzzer randomly mutates the test case
by inserting, deleting, or replacing instructions to generate new
test cases.

The fuzzer runs test cases in two phases: (i) a branch training phase,
with \cc{cond_ptr} storing \cc{true} and \cc{guess_ptr} containing a
correct tag; and (ii) a speculative execution phase, with with
\cc{cond_ptr} storing \cc{false} and \cc{guess_ptr} containing either
a correct or wrong tag.
The fuzzer executes each test case twice.
The first execution runs the branch training phase and then the
speculative execution phase with the correct tag.
The second execution is the same as the first, but the only difference
is to run the speculative execution phase with the wrong tag.
After each execution, the fuzzer measures the access latency of a
cache line and compares the cache state between the two executions.
This process is repeated for each cache line of the 4KB memory
region.
If a notable difference is observed, the fuzzer considers the test
case as a potential MTE tag leakage gadget.

\PP{Results}
We developed the fuzzer and tested it on the same ARMv8.5 devices.
As a result, we additionally identified variants of
\sys-v1~(\autoref{s:gadget1}) that utilize linked list traversal.
The fuzzer was able to discover the gadgets within 1-2 hours of
execution without any prior knowledge of them.

\section{\sys Gadgets}
\label{s:gadget}
In this section, we present \sys gadgets discovered by the tag leakage
template and fuzzing~(\autoref{s:discover}).
Each gadget featuring unique memory access patterns leaks the MTE tag
of a given memory address.
\sys-v1~(\autoref{s:gadget1}) exploits the speculation shrinkage in
branch prediction and data prefetching, and
\sys-v2~(\autoref{s:gadget2}) leverages the blockage of store-to-load
forwarding.
We further analyze the root cause and propose mitigations at hardware
and software levels.

\subsection{\sys-v1: Exploiting Speculation Shrinkage}
\label{s:gadget1}

\begin{figure}[t]
  \centering
  \includegraphics[width=0.75\columnwidth]{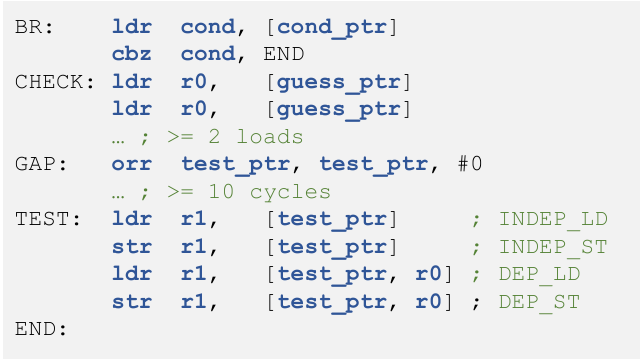}
  \caption{\sys-v1 gadget}
  \label{fig:g1}
\end{figure}

\begin{figure}[t]
  \centering
  \includegraphics[width=\columnwidth]{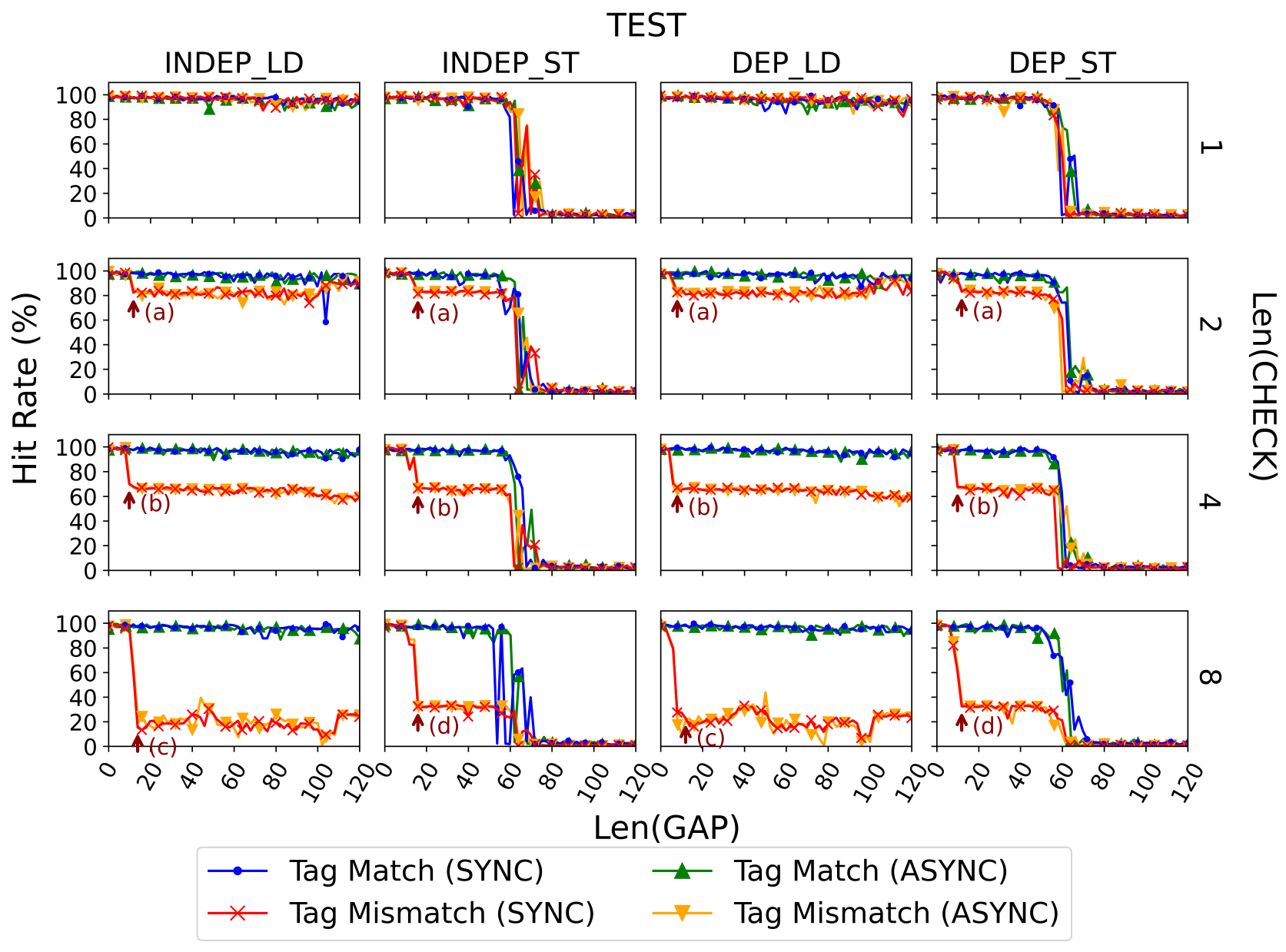}
  \caption{Cache hit rate of \cc{test_ptr} after executing
    \sys-v1. X-axis shows the length of \cc{GAP} and Y-axis shows the
    cache hit rate.}
  \label{fig:g1-graph}
\end{figure}

During our experiment with the MTE tag leakage
template~(\autoref{s:discover:template}), we observed that multiple
tag checks in \cc{CHECK} influence the cache state of \cc{test_ptr}.
With a single tag check, \cc{test_ptr} was always cached regardless of
the tag check result.
However, with two or more tag checks, \cc{test_ptr} was cached on tag
match, but not cached on tag mismatch.

In summary, we found that the following three conditions should hold
for the template to leak the MTE tag:
(i)~\cc{CHECK} should include at least two loads with \cc{guess_ptr};
(ii)~\cc{TEST} should access \cc{test_ptr} (either load or store,
dependent or independent to \cc{CHECK});
and (iii)~\cc{CHECK} should be close to \cc{BR} (within 5 CPU cycles)
while \cc{TEST} should be far from \cc{CHECK} (more than 10 CPU cycles
away).
If these conditions are met, \cc{test_ptr} is cached on tag match and not
cached on tag mismatch.
If any of these is not met, \cc{test_ptr} is either always cached or
never cached.
Based on this observation, we developed \sys-v1, a gadget that leaks
the MTE tag of any given memory address.

\PP{Gadget}
\autoref{fig:g1} illustrates the \sys-v1 gadget. 
In \cc{BR}, the CPU mispredicts the branch result and speculatively
executes \cc{CHECK}.
In \cc{CHECK}, \cc{guess_ptr} is dereferenced two or more times,
triggering tag checks with \cc{Tg} against \cc{Tm}.
\cc{GAP} provides the time gap between \cc{BR} and \cc{TEST}, and then
in \cc{TEST}, \cc{test_ptr} is accessed.
\cc{GAP} can be filled with various types of instructions, such as
computational instructions (e.g., \cc{orr}, \cc{mul}) or memory
instructions (e.g., \cc{ldr}, \cc{str}), as long as it provides more
than 10 CPU cycles of the time gap.

\PP{Experimental Results}
We found that \sys-v1 is an effective MTE tag leakage gadget on the
ARM Cortex-X3 (core 8 of Pixel 8 devices) in both MTE synchronous and
asynchronous modes.
We leveraged the physical CPU cycle counter (i.e., \cc{PMCCNTR_EL0})
for time measurement, which was enabled by modifying the Linux kernel.
An L1 cache hit is determined if the access latency is less than or
equal to 35 CPU cycles.
In a real-world setting, the virtual CPU cycle counter (i.e.,
\cc{CNTVCT_EL0}) is available to the user space with lower resolution,
which also can effectively observe the tag leakage behavior.

We experimented to measure the cache hit rate of \cc{test_ptr} after
executing \sys-v1.
To verify condition~(i), we varied the number of loads in
\cc{CHECK} (i.e., \cc{Len(CHECK)}) from 1, 2, 4, and 8.
To verify condition~(ii), we varied the types of memory access 
in \cc{TEST} (i.e., independent/dependent, load/store).
To verify condition~(iii), we filled \cc{GAP} with a sequence of
\cc{orr} instructions (where each \cc{orr} is dependent on the
previous one) and varied its length (i.e., \cc{Len(GAP)}).
\autoref{fig:g1-graph} shows the experimental results.
The x-axis represents \cc{Len(GAP)} and the y-axis represents the
cache hit rate of \cc{test_ptr} measured over 1,000 trials.

When \cc{Len(CHECK)} was 1, the cache hit rate of \cc{test_ptr} had no
difference between tag match and mismatch.
\cc{test_ptr} was either always cached~(i.e., load access) or cached
until a certain threshold and then not cached~(i.e., store access).
When \cc{Len(CHECK)} was 2 or more, depending on the tag check result,
the cache hit rate differed, validating the condition~(i).
If the tag matched, the cache hit rate was the same as when
\cc{Len(CHECK)} was 1.
If the tag mismatched, the cache hit rate dropped compared to the tag
match~(annotated with arrows).
This difference was observed in all access types of \cc{TEST},
validating the condition~(ii).
The cache hit rate drop was observed after about 10 \cc{orr}
instructions in \cc{GAP}.
Further experiments inserting lengthy instructions between \cc{BR} and
\cc{CHECK}~(\autoref{s:appendix:g1-gap}) confirmed that \cc{CHECK}
should be close to \cc{BR}, validating the condition~(iii).

When the tag mismatched, the cache hit was periodic and the period got
shorter as \cc{Len(CHECK)} increased.
When \cc{Len(CHECK)} is 2, a cache hit occurred in 5 out of 6
trials~(83\%, arrows \cc{(a)}).
When \cc{Len(CHECK)} is 4, a cache hit occurred in 2 out of 3
trials~(66\%, arrows \cc{b}).
When \cc{Len(CHECK)} is 8, the pattern differed between load and store
accesses~(arrows \cc{c} and \cc{d}).
For store access, a cache hit was observed every 3 trials~(33\%, arrows
\cc{c}).
For load access, the pattern changed over iterations from all cache
hits~(0\%) to a cache hit every 2 trials~(50\%)~(arrows \cc{d}).

A similar cache hit rate drop was observed when \cc{guess_ptr} points
to an unmapped address and generated speculative address translation
faults.
This further indicates that \sys-v1 can be utilized as an
address-probing gadget~\cite{blindside} useful for breaking ASLR.

\begin{figure}
  \centering
  \includegraphics[width=\columnwidth]{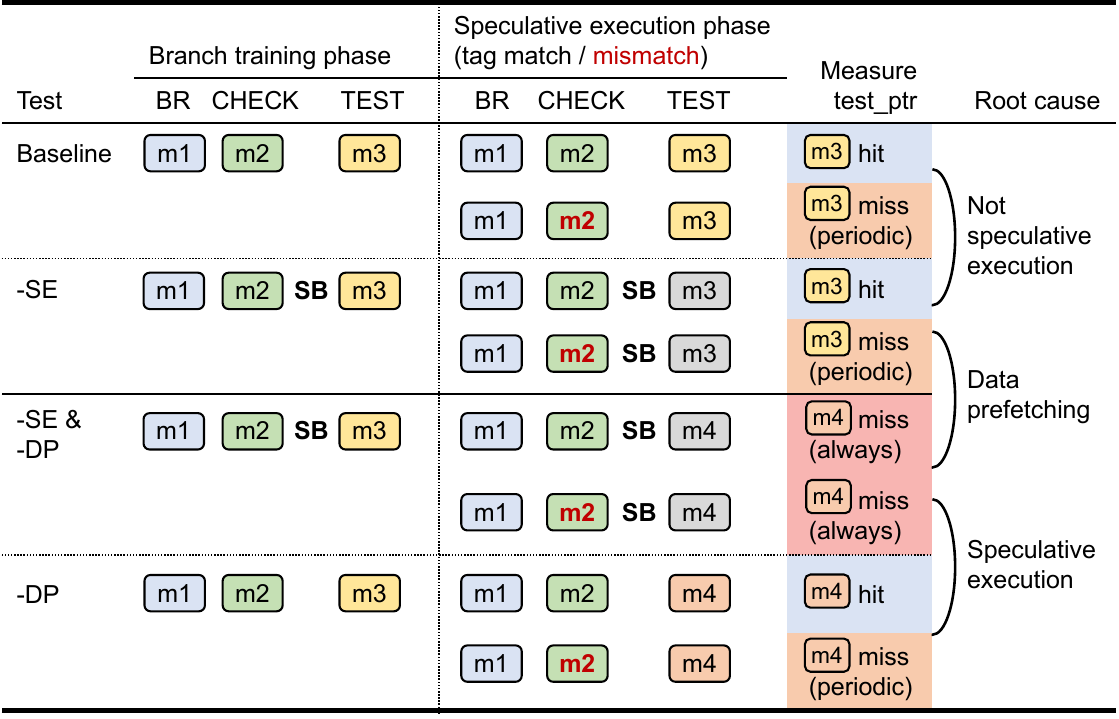}
  \caption{\sys-v1 ablation study. \cc{SE}: Speculative Execution,
    \cc{DP}: Data Prefetch. \cc{m1} and \cc{m2} are memory addresses
    for \cc{cond_ptr} and \cc{target_addr}, respectively. \cc{m3} and
    \cc{m4} are memory address for \cc{test_ptr}. The tests were
    conducted with \cc{LEN(CHECK)=2} and \cc{TEST=INDEP_LD}}
  \label{t:g1-ablation}
\end{figure}

\PP{Root Cause}
Analyzing the gadget, we found that tag check results affect the CPU's
data prefetching behavior and the speculative execution.
This refutes the previous studies on speculative MTE tag
leakage~\cite{pz-mte,sticky-tags}, which stated that tag check faults
do not affect the speculation execution and did not state the impacts
of the data prefetching.

In general, modern CPUs speculatively access memory in two cases:
speculative execution~\cite{spectre} and data
prefetching~\cite{stride,domino}.
To identify the root cause of \sys-v2 in these two cases, we conducted
an ablation study~(\autoref{t:g1-ablation}).
First, we eliminated the effect of speculative execution by inserting
a speculation barrier (i.e., \cc{sb}) between \cc{CHECK} and
\cc{TEST}.
Second, we varied the memory access pattern between branch training
and speculative execution phases to eliminate the effect of data
prefetching.

In \cc{Baseline}, no speculation barrier was inserted, and both branch
training and speculative execution phases accessed the same addresses
in order.
In this case, \cc{test_ptr} was cached on tag match, but not cached
on tag mismatch.
In \cc{-SE}, a speculation barrier was inserted to prevent \cc{TEST}
from being speculatively executed.
Here, the same cache state difference was observed, indicating that
the difference in \cc{Baseline} is not due to the speculative
execution at least in this case.

Next, in \cc{-SE \& -DP}, the memory access pattern was also varied in
the speculative execution phase to prevent \cc{test_ptr} from being
prefetched.
As a result, \cc{test_ptr} was always not cached, verifying that the
CPU failed to prefetch \cc{test_ptr} due to the divergence in the
access pattern.
If we compare \cc{-SE} and \cc{-SE \& -DP}, the cache state difference
was observed only when data prefetching is enabled~(\cc{-SE}).
Considering the CPU's mechanism, such a difference seems to be due to
data prefetching---i.e., the CPU prefetches data based on the previous
access pattern, but skips it on tag mismatch.

Finally, in \cc{-DP}, we removed the speculation barrier to re-enable
the speculative execution of \cc{TEST} while still varying the memory
access pattern.
In this case, the difference is observed again between tag match and
mismatch.
Comparing \cc{-DP} and \cc{-SE \& -DP}, we can conclude that the
speculative execution is also the root cause of the cache
difference---i.e., the CPU halts speculative execution on tag check
faults.

We suspect that the CPU optimizes performance by halting speculative
execution and data prefetching on tag check faults.
A relevant patent filed by ARM~\cite{arm-prefetch} explains that the
CPU can reduce speculations on \emph{wrong path
events}~\cite{wrong-path}, which are events indicating the
possibility of branch misprediction, such as spurious invalid memory
accesses.
By detecting branch misprediction earlier, the CPU can save recovery
time from wrong speculative execution and improve the data prefetch
accuracy by not prefetching the wrong path-related data.
Since these optimizations are beneficial in both MTE synchronous and
asynchronous modes, we think that the tag leakage behaviors were
observed in both MTE modes.

We also think there is a time window to detect wrong path events
during speculative execution upon branch prediction, which seems to be
5 CPU cycles.
As explained in the patent~\cite{arm-prefetch}, the CPU may maintain
speculation confidence values for speculative execution and data
prefetching.
We think the CPU reduces the confidence values on tag check faults,
halts speculation if it drops below a certain threshold, and restores
it to the initial level.
This reasoning explains the periodic cache miss of \cc{test_ptr} on
tag mismatch, where the confidence value is repeatedly reduced below
the threshold~(i.e., cache miss) and restored~(i.e., cache hit).
\JH{TODO: example?}
In addition, when \cc{TEST} is store access, the speculation barrier
made \cc{test_ptr} always not cached.
This indicates that the CPU does not prefetch data for store access,
thus the speculation window shrinkage is the only root cause in such 
cases.

\begin{figure}[t]
  \centering
  \begin{subfigure}{0.47\columnwidth}
    \centering
    \includegraphics[width=\columnwidth]{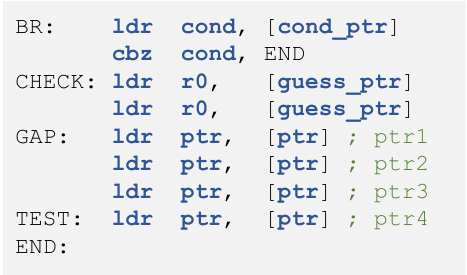}
    \caption{variant 1}
    \label{fig:g1-link-1}
  \end{subfigure}
  \begin{subfigure}{0.47\columnwidth}
    \centering
    \includegraphics[width=\columnwidth]{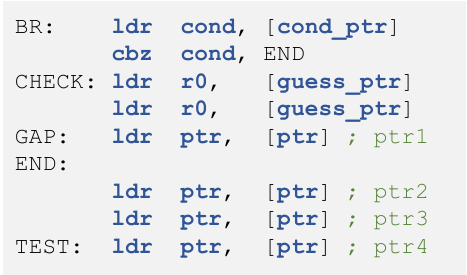}
    \caption{variant 2}
    \label{fig:g1-link-2}
  \end{subfigure}
  \caption{\sys-v1 gadget variants}
  \label{fig:g1-var}
\end{figure}

\PP{Variants} 
Running the tag leakage fuzzer~(\autoref{s:discover:fuzzing}), we
discovered variants of \sys-v1 leveraging linked list
traversal~(\autoref{fig:g1-var}).
Before the gadget, a linked list of 4 instances is initialized, where
each instance points to the next instance (i.e., \cc{ptr0} to
\cc{ptr3}), and the cache line of the last instance~(\cc{ptr3}) is
flushed.
The gadget traverses the linked list by accessing \cc{ptr0} to
\cc{ptr3} in order, where \cc{TEST} accesses \cc{ptr3} only if the
branch result is true.
After the gadget, the cache hit rate of \cc{ptr3} is measured.

In the first variant~(\autoref{fig:g1-link-1}), \cc{TEST} is located
in the true branch of the conditional branch \cc{BR}.
In the second variant~(\autoref{fig:g1-link-2}), \cc{TEST} is located
out of the conditional branch path, where both the true and false
branches merge.
In both variants, if the tag matched in \cc{CHECK}, \cc{ptr3} was
cached, but not cached if the tag mismatched, as in the original
\sys-v1 gadget~(\autoref{fig:g1-graph}).
We think the root cause is the same as the original gadget---i.e., the
CPU changes speculative execution and data prefetching behaviors on tag
check faults.
Moreover, the variants can be effective in realistic scenarios like
linked list traversal, and the tag leakage can also be observed from
the memory access outside the conditional branch's scope.

\PP{Mitigation}
\sys-v1 exploits the speculation shrinkage on tag check faults in
speculative execution and data prefetching.
To prevent MTE tag leakage at the micro-architectural level, the CPU
should not change the speculative execution or data prefetching
behavior on tag check faults.

To prevent \sys-v1 at the software level, the following two approaches
can be used.

i) \emph{Speculation barrier}:
Speculation barrier can prevent the \sys-v1 gadget from leaking the
MTE tag.
If \cc{TEST} contains store access, placing a speculation barrier
(i.e., \cc{sb}) or instruction synchronization barrier (i.e.,
\cc{isb}) makes \cc{test_ptr} always not cached, preventing the tag
leakage.
If \cc{TEST} contains load access, placing the barrier before
\cc{TEST} does not prevent tag leakage, but placing it before
\cc{CHECK} mitigates tag leakage, since speculative tag check faults
are not raised.

ii) \emph{Padding instructions}:
\sys-v1 requires that tag check faults are raised within a time window
from the branch.
By inserting a sequence of instructions before \cc{CHECK} to extend
the time window, the CPU does not reduce the speculations and
\cc{test_ptr} is always cached~(\autoref{s:appendix:g1-gap}).

\subsection{\sys-v2: Exploiting Store-to-Load Forwarding}
\label{s:gadget2}

\begin{figure}[t]
    \centering
    \includegraphics[width=0.75\columnwidth]{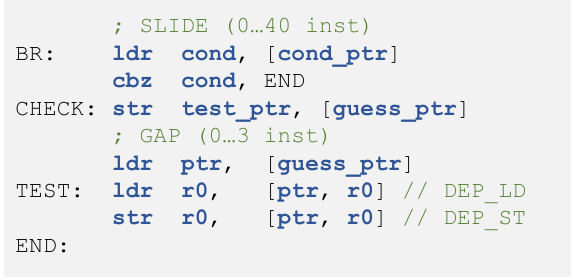}
    \caption{\sys-v2: A speculative tag check fault blocks
      store-to-load forwarding}
    \label{fig:g2}
\end{figure}

\begin{figure}[t]
  \centering
  \includegraphics[width=0.9\columnwidth]{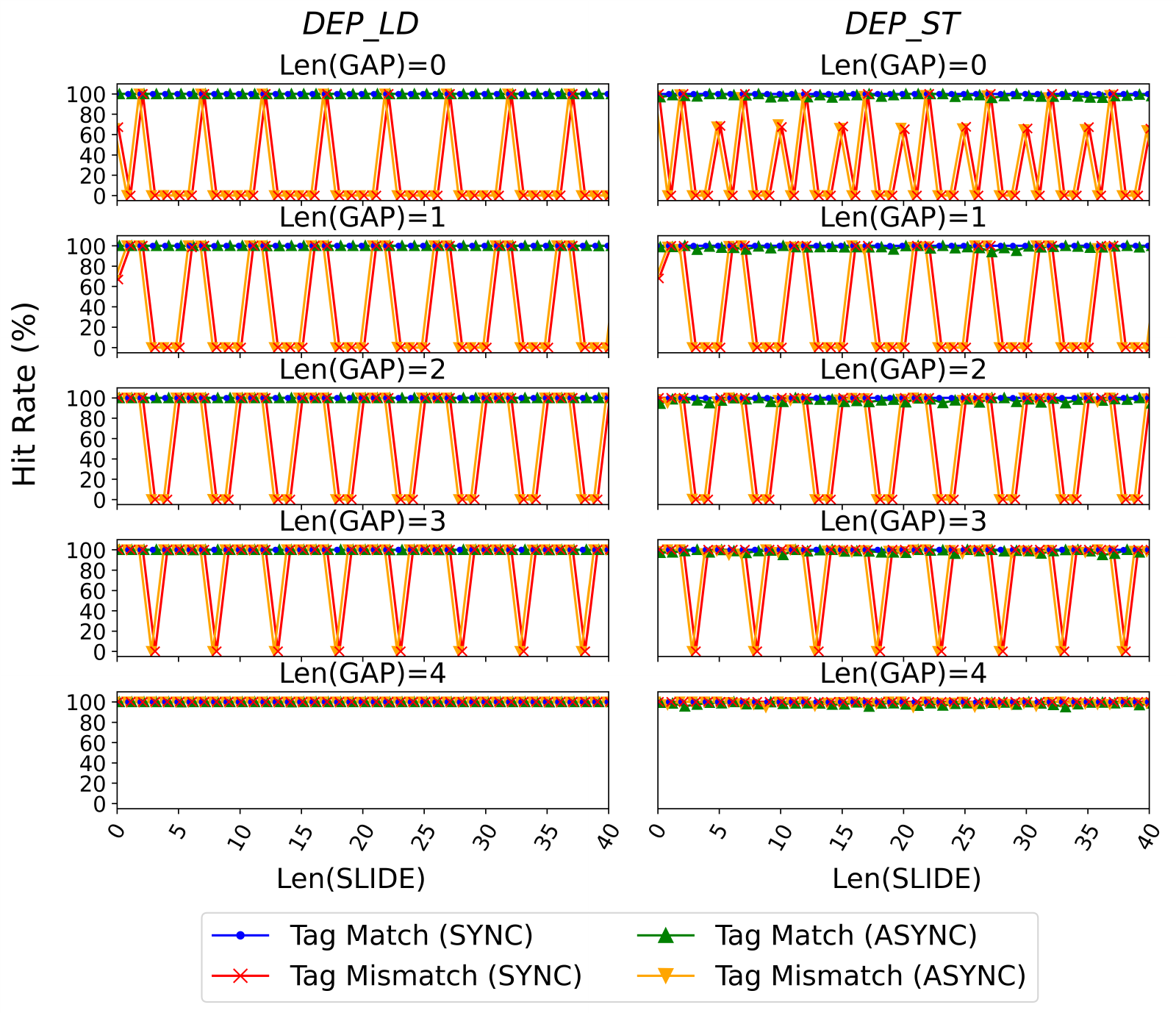}
  \caption{Cache hit rate of \cc{test_addr} after executing \sys-v2}
  \label{fig:g2-graph}
\end{figure}

Inspired by Spectre-v4~\cite{spectre-v4} and LVI~\cite{lvi} attacks,
we experimented with the MTE tag leakage template to trigger
store-to-load forwarding behavior~\cite{store-load}.
As a result, we discovered that store-to-load forwarding behavior
differs on tag check result if the following conditions hold:
(i)~\cc{CHECK} triggers store-to-load forwarding, and (ii)~\cc{TEST}
accesses memory dependent on the forwarded value.
If the tag matches in \cc{CHECK}, the memory accessed in \cc{TEST} is
cached; otherwise, it is not cached.
Based on this observation, we developed \sys-v2.

\PP{Gadget}
\autoref{fig:g2} illustrates the \sys-v2 gadget.
The initial setting follows the MTE tag leakage
template~(\autoref{s:discover:template}), where the CPU mispredicts
the branch \cc{BR} and speculatively executes \cc{CHECK} and
\cc{TEST}.
At \cc{CHECK}, the CPU triggers store-to-load forwarding behavior by
storing \cc{test_ptr} at \cc{guess_ptr} and immediately loading the
value from \cc{guess_ptr} as \cc{ptr}.
If \cc{CHECK} succeeds the tag checks (i.e., \cc{Tg} matches \cc{Tm}),
the CPU forwards \cc{test_ptr} to \cc{ptr} and speculatively accesses
\cc{test_ptr} in \cc{TEST}.
If \cc{CHECK} fails the tag checks (i.e., \cc{Tg} mismatches \cc{Tm}),
the CPU blocks \cc{test_ptr} from being forwarded to \cc{ptr} and
does not access \cc{test_ptr}.

\PP{Experimental Results}
\sys-v2 showed its effectiveness as an MTE tag leakage gadget in ARM
Cortex-A715 (core 4-7 of Pixel 8).
We identified one requirement for \sys-v2 to exhibit the tag leakage
behavior: the store and load instructions in \cc{CHECK} should be
executed within 5 instructions.
If this requirement is met, the cache hit rate of \cc{test_ptr} after
\sys-v2 exhibited a notable difference between tag match and
mismatch~(\autoref{fig:g2-graph}).
Otherwise, the store-to-load forwarding always succeeded and the CPU
forwarded \cc{test_ptr} to \cc{ptr}, and \cc{test_ptr} was always
cached.

To verify the requirement, we conducted experiments by inserting two
instruction sequences, \cc{SLIDE} and \cc{GAP}~(\autoref{fig:g2}).
Both sequences consist of bitwise OR operations (\cc{orr}), each
dependent on the previous one, while not changing register or memory
states.
\cc{SLIDE} is added before the gadget to control the alignment of
store-to-load forwarding in the CPU pipeline.
\cc{GAP} is added between the store and load instructions in
\cc{CHECK} to control the distance between them.
The results varying \cc{SLIDE} from 0 to 40 and \cc{GAP} from 0 to 4
are shown in \autoref{fig:g2-graph}.
In each subfigure, the x-axis represents the length of \cc{GAP},
and the y-axis represents the cache hit rate of \cc{test_ptr} after
the gadget, measured over 1000 trials.

On tag match, the cache hit rate of \cc{test_ptr} was near 100\% in all
conditions.
However, on tag mismatch, the hit rate dropped on every 5 instructions
in \cc{SLIDE}.
With \cc{Len(GAP)=0} (i.e., no instruction), the hit rate drop was
most frequent, occurring in 4 out of \cc{SLIDE} length.
With \cc{Len(GAP)=1,2,3}, the hit rate drop occurred in 3, 2, and 1
times every 5 instructions in \cc{SLIDE}, respectively.
With \cc{Len(GAP)=4}, no hit rate drop was observed.

Similarly to \sys-v1, the blockage of store-to-load forwarding was not
specific to the MTE tag check fault, but was also observed with
address translation fault, thus \sys-v2 can also be utilized as an
address-probing gadget.

\PP{Root Cause}
The root cause of \sys-v2 is likely due to the CPU preventing
store-to-load forwarding on tag check faults.
The CPU detects the store-to-load dependency utilizing internal
buffers that log memory access information, such as Load-Store
Queue~(LSQ), and forwards the data if the dependency is detected.
Although there is no documentation detailing the store-to-load
forwarding mechanism on tag check faults, a relevant patent filed by
ARM~\cite{arm-check-skip} provides a hint on the possible explanation.

The patent suggests that if the store-to-load dependency is detected,
the load instruction can skip the tag check and the CPU can always
forward the data.
If so, store-to-load forwarding would not leak the tag check result
(i.e., data is forwarded both on tag match and mismatch), as observed
when \cc{Len(GAP)} is 4 or more.
When \cc{Len(GAP)} is less than 4, however, the store-to-load
succeeded on tag match and failed on tag mismatch.

We suspect that the CPU performs the tag check for the load
instruction if the store-to-load dependency is not detected, and the
CPU blocks the forwarding on tag check faults to prevent meltdown-like
attacks~\cite{meltdown,fallout}.
Considering the affected core~(i.e., Cortex-A715) dispatches 5
instructions in a cycle~\cite{a715-slides}, it is likely that the CPU
cannot detect the dependency if the store and load instructions are
executed in the same cycle, since the store information is not yet
written to the internal buffers.
If \cc{Len(GAP)} is 4 or more, the store and load instructions are
executed in the different cycles, and the CPU can detect the
dependency.
Therefore, the CPU skips the tag check and always forwards the data
from the store to load instructions.
If \cc{Len(GAP)} is less than 4, the store and load instructions are
executed in the same cycle, and the CPU fails to detect the dependency
and performs the tag check for the load instruction.
In this case, the forwarding is blocked on tag check faults.

\PP{Mitigation}
To prevent tag leakage in \sys-v2 at the micro-architectural level,
the CPU should be designed to either always allow or always block the
store-to-load forwarding regardless of the tag check result.
Always blocking the store-to-load forwarding may raise a performance
issue.
Instead, always allowing forwarding can effectively prevent the tag
leakage with low-performance overheads.
This would not introduce meltdown-like vulnerabilities, because tag
mismatch occurs within the same exception level.

At the software level, the following mitigations can be applied to
prevent \sys-v2.

i) \emph{Speculation barrier}: If a speculation barrier is inserted
before \cc{CHECK}, the CPU does not speculatively execute
store-to-load forwarding on both tag match and mismatch.
Thus, \cc{test_ptr} is not cached regardless of the tag check result.

ii) \emph{Preventing Gadget Construction}: If the store and load
instructions in \cc{CHECK} are not executed within 5 instructions, the
store-to-load forwarding is always allowed on both cases, making
\cc{test_ptr} cached always.
The potential gadgets can be modified to have more than 5 instructions
between the store and load instructions in \cc{CHECK}, by adding dummy
instructions (e.g., \cc{nop}) or reordering the instructions.

\section{Real-World Attacks}
\label{s:attack}

To demonstrate the exploitability of \sys gadgets in MTE-based
mitigation, this section develops two real-world attacks against
Chrome and Linux kernel~(\autoref{fig:attacks}).
There are several challenges to launching real-world attacks using
\sys gadgets.
First, \sys gadgets should be executed in the target address space,
requiring the attacker to construct or find gadgets from the target
system.
Second, the attacker should control and observe the cache state to
leak the tag check results.
In the following, we demonstrate the real-world attacks using \sys
gadgets on two real-world systems: the Google Chrome
browser~(\autoref{s:chrome}) and the Linux kernel~(\autoref{s:linux}),
and discuss the mitigation strategies.

\subsection{Attacking Chrome Browser}
\label{s:chrome}
A web browser is a primary attack surface for web-based attacks as it
processes untrusted web content, such as JavaScript and HTML.
We first overview the threat model~(\autoref{s:chrome-model}) and
provide a \sys gadget constructed in the V8 JavaScript
engine~(\autoref{s:chrome-gadget}).
Then, we demonstrate the effectiveness of \sys gadgets in exploiting
the browser~(\autoref{s:chrome-exploit}) and discuss the mitigation
strategies~(\autoref{s:chrome-mitigation}).

\subsubsection{Threat Model}
\label{s:chrome-model}
We follow the typical threat model of Chrome browser attacks, where
the attacker aims to exploit memory corruption vulnerabilities in the
renderer process.
We assume the victim user visits the attacker-controlled website,
which serves a malicious webpage.
The webpage includes crafted HTML and JavaScript, which exploit memory
corruption vulnerabilities in the victim's renderer process.
We assume all Chrome's state-of-the-art mitigation techniques are in
place, including ASLR~\cite{aslr}, CFI~\cite{chromium-cfi}, site
isolation~\cite{site-isolation}, and V8 sandbox~\cite{v8-sandbox}.
Additionally, as an orthogonal defense, we assume that the renderer
process enables random MTE tagging in PartitionAlloc~\cite{partition}.

\begin{figure}[t]
  \centering
  \begin{subfigure}{0.8\columnwidth}
    \centering
    \include{code/g2.js}
    \vspace{-1.5em}
    \caption{JavaScript \sys-v2 gadget}
    \label{code:js-gadget}
  \end{subfigure}
  \begin{subfigure}{0.8\columnwidth}
    \centering
    \include{code/g2-js.c}
    \vspace{-1.5em}
    \caption{Pseudo C code of JIT-optimized JavaScript \sys-v2 gadget}
    \label{code:js-c}
  \end{subfigure}
  \caption{V8 \sys-v2 gadget in JavaScript}
  \label{code:js-v2}
\end{figure}

\subsubsection{Constructing \sys Gadget}
\label{s:chrome-gadget}

In the V8 JavaScript environment, \sys-v2 was successfully constructed
and leaked the MTE tags of any memory address. 
However, we didn't find a constructible \sys-v1 gadget, since the
tight timing constraint between \cc{BR} and \cc{CHECK} was not
feasible in our speculative V8 sandbox escape
technique~(\autoref{s:appendix:v8-sandbox}).

\PP{V8 TikTag-v2 Gadget}
\autoref{code:js-v2} is the \sys-v2 gadget constructed in the V8
JavaScript engine and its pseudo-C code after JIT compilation.
With this gadget, the attacker can learn whether the guessed tag \cc{Tg}
matches with the tag \cc{Tm} assigned to \cc{target_addr}.
The attacker prepares three arrays, \cc{slow}, \cc{victim},
\cc{probe}, and an \cc{idx} value.
\cc{slow} is a \cc{Unit8Array} with a length of 64 and is accessed in
\cc{BR} to trigger the branch misprediction.
\cc{victim} is a \cc{Float64Array} with length 64, which is accessed
to trigger store-to-load forwarding.
\cc{probe} is a \cc{Uint8Array} with length 512, and is accessed in
\cc{TEST} to leak the tag check result.
A \cc{Number} type \cc{idx} value is used in out-of-bounds access of
\cc{victim}.
\cc{idx} value is chosen such that \cc{victim[idx]} points to
\cc{target_addr} with a guessed tag \cc{Tg} (i.e.,
\cc{(Tg<<56)|target_addr}).
To speculatively access the \cc{target_addr} outside the V8 sandbox,
we leveraged the speculative V8 sandbox escape technique we discovered
during our research, which we detail
in~\autoref{s:appendix:v8-sandbox}.

Line 8 of~\autoref{code:js-gadget} is the \cc{BR} block of the \sys-v2
gadget, triggering branch misprediction with \cc{slow[0]}.
Line 12-13 is the \cc{CHECK} block, which performs the store-to-load
forwarding with \cc{victim[idx]}, accessing \cc{target_addr} with a
guessed tag \cc{Tg}.
When this code is JIT-compiled~(\autoref{code:js-c}), a bound check is
performed, comparing \cc{idx} against \cc{victim.length}.
If \cc{idx} is an out-of-bounds index, the code returns
\cc{undefined}, but if \cc{victim.length} field takes a long time to
be loaded, the CPU speculatively executes the following store and load
instructions.
After that, line 17 implements the \cc{TEST} block, which accesses the
\cc{probe} with the forwarded value \cc{val} as an index.
Again, a bound check on \cc{val} against the length of \cc{probe} is
preceded, but this check succeeds as \cc{PROBE_OFFSET} is smaller than
the length of \cc{probe} array.
As a result, \cc{probe[PROBE_OFFSET]} is cached only when the
store-to-load forwarding succeeds, which is the case when \cc{Tg}
matches \cc{Tm}.

\begin{figure}[t]
  \centering
  \begin{subfigure}{0.38\columnwidth}
    \centering
    \includegraphics[width=\columnwidth]{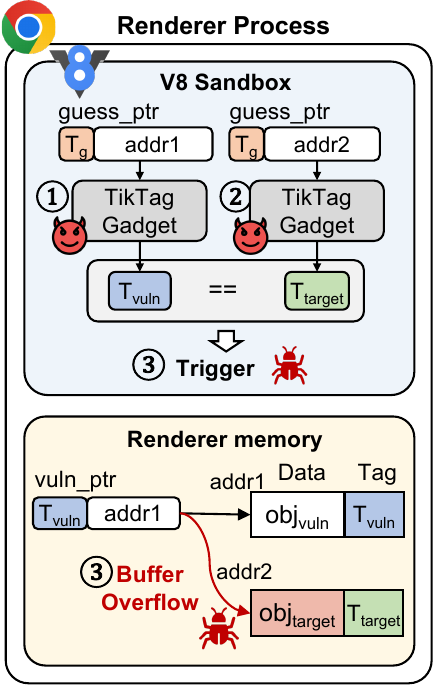}
    \caption{Chrome browser}
    \label{fig:chrome-attack}
  \end{subfigure}
  \hspace{1em}
  \begin{subfigure}{0.5\columnwidth}
    \centering
    \includegraphics[width=\columnwidth]{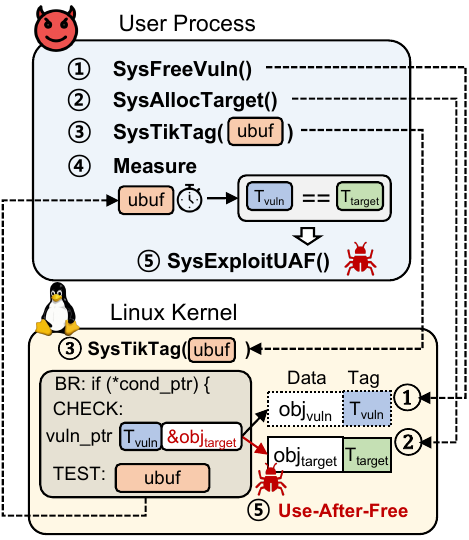}
    \caption{Linux kernel}
    \label{fig:kernel-attack}
  \end{subfigure}
  \caption{MTE bypass attacks}
\label{fig:attacks}
\end{figure}

\subsubsection{Chrome MTE bypass attack}
\label{s:chrome-exploit}
\autoref{fig:chrome-attack} illustrates the overall MTE bypass attack
on the Chrome browser with arbitrary tag leakage primitive of \sys
gadgets.
We assume a buffer overflow vulnerability in the renderer process,
where exploiting a temporal vulnerability (e.g., use-after-free) is
largely the same.
The vulnerability overflows a pointer (i.e., \cc{vuln_ptr}) to a
vulnerable object (i.e., \objvuln), corrupting the adjacent object
(i.e., \objtarget).
With PartitionAlloc's MTE enforcement, two objects have different tags
with a 14/15 probability.
To avoid raising an exception, the attacker needs to ensure that the
tags of \objvuln and \objtarget are the same.
\sys-v2 can be utilized to leak the tag of \objvuln~(\C{1}) and
\objtarget~(\C{2}).
If both leaked tags are the same, the attacker exploits the
vulnerability, which would not raise a tag check fault~(\C{3}).
Otherwise, the attacker frees and re-allocates \objtarget and goes
back to the first step until the tags match.

\PP{Triggering Cache Side-Channel}
To successfully exploit a \sys gadget, the attacker needs to satisfy
the following requirements: i) branch training, ii) cache control,
and iii) cache measurement.
All three requirements can be met in JavaScript.
First, the attacker can train the branch predictor by running the
gadget with non-zero \cc{slow[0]} and in-bounds \cc{idx}, and trigger
the branch misprediction in \cc{BR} with zero value in \cc{slow[0]}
and out-of-bounds \cc{idx}.
Second, the attacker can evict the cache lines of \cc{slow[0]},
\cc{victim.length}, and \cc{probe[PROBE_OFFSET]} with JavaScript
cache eviction techniques~\cite{evset, leaky-page, spookjs}.
Third, the attacker can measure the cache status of
\cc{probe[PROBE_OFFSET]} with a high-resolution timer based on
\cc{SharedArrayBuffer}~\cite{SharedArrayBuffer, fantastic}.

\PP{Exploiting Memory Corruption Vulnerabilities}
Given the leaked MTE tags, the attacker can exploit spatial and
temporal memory corruption vulnerabilities in the renderer.
The attack strategy is largely the same as the traditional memory
corruption attacks but should ensure that the vulnerability does not
raise a tag check fault utilizing the leaked tags.
We further detail the attack strategy
in~\autoref{s:appendix:v8-exploit}.

\subsubsection{Mitigation}
\label{s:chrome-mitigation}
To mitigate the \sys gadget-based MTE bypass attacks in the browser
renderer process, the following mitigations can be employed:

i) \emph{Speculative execution-aware sandbox}:
To stop attackers from launching \sys-based attacks from a sandboxed
environment like V8 sandbox, the sandbox can be fortified by
preventing any speculative memory access beyond the sandbox's memory
region.
While modern web browsers employ a sandbox to isolate untrusted web
contents from the renderer, they often overlook speculative paths.
For instance, Chrome V8 sandbox~\cite{v8-sandbox} and Safari Webkit
sandbox~\cite{gigacage} do not completely mediate the speculative
paths~\cite{ileakage}.
Based on current pointer compression
techniques~\cite{pointer-compression}, speculative paths can be
restricted to the sandbox region by masking out the high bits of the
pointers.

ii) \emph{Speculation barrier}:
As suggested in~\autoref{s:gadget}, placing a speculation barrier
after \cc{BR} for potential \sys gadgets can prevent speculative tag
leakage attacks.
However, this mitigation may not be applicable in the
performance-critical browser environment, as it may introduce
significant performance overhead.

iii) \emph{Prevention of gadget construction}:
As suggested in~\autoref{s:gadget2}, the \sys-v2 gadget can be
mitigated by padding instructions between store and load instructions.
A \sys-v1 gadget, although we have not found an exploitable one, can
be mitigated by padding instructions between a branch and memory
accesses, as described in~\autoref{s:gadget1}.

\subsection{Attacking the Linux Kernel}
\label{s:linux}
The Linux kernel on ARM is widely used for mobile devices, servers,
and IoT devices, making it an attractive attack target.
Exploiting a memory corruption vulnerability in the kernel can
escalate the user's privilege, and thus MTE is a promising protection
mechanism for the Linux kernel.
\sys-based attacks against the Linux kernel pose unique challenges
different from the browser attack~(\autoref{s:chrome}).
This is because the attacker's address space is isolated from the
kernel's address space where the gadget will be executed.
In the following, we first overview the threat model of the Linux
kernel~(\autoref{s:linux-model}) and provide a proof-of-concept \sys
gadget we discovered in the Linux kernel~(\autoref{s:linux-gadget}).
Finally, we demonstrate the effectiveness of \sys gadgets in
exploiting Linux kernel vulnerabilities~(\autoref{s:linux-exploit}).

\subsubsection{Threat Model}
\label{s:linux-model}
The threat model here is largely the same as that of typical privilege
escalation attacks against the kernel.
Specifically, we focus on the ARM-based Android Linux kernel, hardened
with default kernel protections (e.g., KASLR, SMEP, SMAP, and CFI).
We further assume the kernel is hardened with an MTE random tagging
solution, similar to the production-ready MTE solutions,
Scudo~\cite{scudo}.
To be specific, each memory object is randomly tagged, and a random
tag is assigned when an object is freed, thereby preventing both
spatial and temporal memory corruptions.

The attacker is capable of running an unprivileged process and aims
to escalate their privilege by exploiting memory corruption
vulnerabilities in the kernel.
It is assumed that the attacker knows kernel memory corruption
vulnerabilities but does not know any MTE tag of the kernel memory.
Triggering memory corruption between kernel objects with mismatching
tags would raise a tag check fault, which is undesirable for
real-world exploits.

One critical challenge in this attack is that the gadget should be
constructed by reusing the existing kernel code and executed by the
system calls that the attacker can invoke.
As the ARMv8 architecture separates user and kernel page tables, user
space gadgets cannot speculatively access the kernel memory.
This setup is very different from the threat model of attacking the
browser~(\autoref{s:chrome}), which leveraged the attacker-provided
code to construct the gadget.
We excluded the \cc{eBPF}-based gadget construction
either~\cite{ebpf, spec-type}, because \cc{eBPF} is not available for
the unprivileged Android process~\cite{arm-cache}.

\begin{figure}[t]
  \centering
  \input{code/snd-timer.c}
  \caption{\sys-v1 gadget in snd_timer_user_read(). -/+ denotes the code changes to make the gadget exploitable.}
  \label{code:kernel-gadget}
\end{figure}

\subsubsection{Kernel TikTag Gadget}
\label{s:linux-gadget}

As described in~\autoref{s:discover:template}, \sys gadgets should
meet several requirements, and each requirement entails challenges in
the kernel environment.

First, in \cc{BR}, a branch misprediction should be triggered with
\cc{cond_ptr}, which should be controllable from the user space.
Since recent AArch64 processors isolate branch prediction training
between the user and kernel~\cite{arm-cache}, the branch training
needs to be performed from the kernel space.
Second, in \cc{CHECK}, \cc{guess_ptr} should be dereferenced.
\cc{guess_ptr} should be crafted from the user space such that it
embeds a guess tag (\cc{Tg}) and points to the kernel address (i.e.,
\cc{target_addr}) to leak the tag (\cc{Tm}).
Unlike the browser JavaScript environment~(\autoref{s:chrome}),
user-provided data is heavily sanitized in system calls, so it is
difficult to create an arbitrary kernel pointer.
For instance, \cc{access_ok()} ensures that the user-provided pointer
points to the user space, and the \cc{array_index_nospec} macro
prevents speculative out-of-bounds access with the user-provided
index.
Thus, \cc{guess_ptr} should be an existing kernel pointer,
specifically the vulnerable pointer that causes memory corruption.
For instance, a dangling pointer in use-after-free (UAF) or an
out-of-bounds pointer in buffer overflow can be used.
Lastly, in \cc{TEST}, \cc{test_ptr} should be dereferenced, and
\cc{test_ptr} should be accessible from the user space.
To ease the cache state measurement, \cc{test_ptr} should be a user
space pointer provided through a system call argument.

\PP{Discovered Gadgets}
We manually analyzed the source code of the Linux kernel to find the
\sys gadget meeting the aforementioned requirements.
As a result, we found one potentially exploitable \sys-v1 gadget in
\cc{snd_timer_user_read()}~(\autoref{code:kernel-gadget}).
This gadget fulfills the requirements of
\sys-v1~(\autoref{s:gadget1}).
At line 10 (i.e., \cc{BR}), the \cc{switch} statement triggers branch
misprediction with a user-controllable value \cc{tu->tread} (i.e.,
\cc{cond_ptr}).
At lines 14-17 (i.e., \cc{CHECK}), \cc{tread} (i.e., \cc{guess_ptr})
is dereferenced by four load instructions.
\cc{tread} points to a \cc{struct snd_timer_tread64} object that the
attacker can arbitrarily allocate and free.
If a temporal vulnerability transforms \cc{tread} into a dangling
pointer, it can be used as a \cc{guess_ptr}.
At line 20, (i.e., \cc{TEST}), a user space pointer \cc{buffer}
(i.e., \cc{test_ptr}) is dereferenced in \cc{copy_to_user}.

As this gadget is not directly reachable from the user space, we made
a slight modification to the kernel code; we removed the early return
for the \cc{default} case at line 6.
This ensures that the \cc{buffer} is only accessed in the speculative
path to observe the cache state difference due to speculative
execution.
Although this modification is not realistic in a real-world scenario,
it demonstrates the potential exploitability of the gadget if similar
code changes are made.

We discovered several more potentially exploitable gadgets, but we
were not able to observe the cache state difference between the tag
match and mismatch.
Still, we think there is strong potential for exploiting those
gadgets.
Launching \sys-based attacks involves complex and sensitive
engineering, and thus we were not able to experiment with all possible
cases.
Especially, \sys-v1 relies on the speculation shrinkage on wrong path
events, which may also include address translation faults or other
exceptions in the branch misprediction path.
As system calls involve complex control flows, the speculation
shrinkage may not be triggered as expected.
In addition, several gadgets may become exploitable when kernel code
changes.
For instance, a \sys-v1 gadget in \cc{ip6mr_ioctl()} did not exhibit
an MTE tag leakage behavior when called from its system call path
(i.e., \cc{ioctl}).
However, the gadget had tag leakage when it was ported to other
syscalls (e.g., \cc{write}) with a simple control flow.

\subsubsection{Kernel MTE bypass attack}
\label{s:linux-exploit}

\autoref{fig:kernel-attack} illustrates the MTE bypass attacks on the
Linux kernel.
Taking a use-after-free vulnerability as an example, we assume the
attacker has identified a corresponding \sys gadget,
\cc{SysTikTagUAF()}, capable of leaking the tag check result of the
dangling pointer created by the vulnerability.
For instance, the \sys-v1 gadget in
\cc{snd_timer_user_read()}~(\autoref{code:kernel-gadget}) can leak
the tag check result of \cc{tread}, which can become a dangling
pointer by a use-after-free or double-free vulnerability.

The attack proceeds as follows:
First, the attacker frees a kernel object (i.e., \objvuln) and leaves
its pointer (i.e., \cc{vuln_ptr}) as a dangling pointer~(\C{1}).
Next, the attacker allocates another kernel object (i.e., \objtarget)
at the address of \objvuln with \cc{SysAllocTarget()}~(\C{2}).
Then, the attacker invokes \cc{SysTikTag()} with a user space buffer
(i.e., \cc{ubuf})~(\C{3}), and leaks the tag check result (i.e.,
\cc{Tm} == \cc{Tg}) by measuring the access latency of
\cc{ubuf}~(\C{4}).
If the tags match, the attacker triggers \cc{SysExploitUAF()}, a
system call that exploits the use-after-free vulnerability~(\C{5}).
Otherwise, the attacker re-allocates \objtarget until the tags match.

\PP{Triggering Cache Side-Channel}
As in~\autoref{s:chrome-exploit}, a successful \sys gadget
exploitation requires i) branch training, ii) cache control, and iii)
cache measurement.
For branch training, the attacker can train the branch predictor and
trigger speculation with user-controlled branch conditions from the
user space.
For cache control, the attacker can flush the user space buffer
(i.e., \cc{ubuf}), while the kernel memory address can be evicted by
cache line bouncing~\cite{cache-line-bouncing}.
For cache measurement, the access latency of \cc{ubuf} can be measured
with the virtual counter (i.e., \cc{CNTVCT_EL0}) or a memory
counter-based timer (i.e., near CPU cycle resolution).

\PP{Exploiting Memory Corruption Vulnerabilities}
\sys gadgets enable bypassing MTE and exploiting kernel memory
corruption vulnerabilities.
The attacker can invoke the \sys gadget in the kernel to speculatively
trigger the memory corruption and obtain the tag check result.
Then, the attacker can obtain the tag check result, and trigger the
memory corruption only if the tags match.
We detail the Linux kernel MTE bypass attack process
in~\autoref{s:appendix:linux}.

\subsubsection{Mitigation}
\label{s:linux-mitigation}
To mitigate \sys gadget in the Linux kernel, the kernel developers
should consider the following mitigations:

i) \emph{Speculation barrier}:
Speculation barriers can effectively mitigate \sys-v1 gadget in the
Linux kernel.
To prevent attackers from leaking the tag check result through the
user space buffer, kernel functions that access user space addresses,
such as \cc{copy_to_user} and \cc{copy_from_user}, can be hardened
with speculation barriers.
As described in~\autoref{s:gadget1}, leaking tag check results with
store access can be mitigated by placing a speculation barrier before
the store access (i.e., \cc{TEST}).
For instance, to mitigate the gadgets leveraging \cc{copy_to_user},
a speculation barrier can be inserted before the \cc{copy_to_user}
invocation.
For gadgets utilizing load access to the user space buffer, the 
barriers mitigate the gadgets if inserted between the branch and the 
kernel memory access (i.e., \cc{CHECK}).
For instance, to mitigate the gadgets leveraging \cc{copy_from_user},
the kernel developers should carefully analyze the kernel code base to
find the pattern of the conditional branch, kernel memory access, and
\cc{copy_from_user()}, and insert a speculation barrier between the
branch and the kernel memory access.

ii) \emph{Prevention of gadget construction}:
To eliminate potential \sys gadgets in the Linux kernel, the kernel
source code can be analyzed and patched.
As \sys gadgets can also be constructed by compiler optimizations, a 
binary analysis can be conducted.
For each discovered gadget, instructions can be reordered or
additional instructions can be inserted to prevent the gadget
construction, following the mitigation strategies
in~\autoref{s:gadget1} and~\autoref{s:gadget2}.

\section{Evaluation}
\label{s:eval}

In this section, we evaluate the \sys gadgets and MTE bypass exploits
in two MTE-enabled systems, the Chrome
browser~(\autoref{s:eval:chrome}) and the Linux
kernel~(\autoref{s:eval:linux}).
All experiments were conducted on the Google Pixel 8 devices.

\subsection{Chrome Browser Tag Leakage}
\label{s:eval:chrome}

\begin{table}
    \centering
    \scriptsize
    \caption{MTE schemes in Android and Chrome allocators.}
    \label{t:allocator}
    \begin{threeparttable}
        \begin{tabular}{l|l|l}
            \toprule
            \textbf{Allocator} & \textbf{Scudo} & \textbf{PartitionAlloc} \\
            \midrule
            \textbf{Usage} & Android & Chrome \\
            \textbf{Allocation (new)} & Random (0x0-0xf)* & Random (0x1-0xf) \\
            \textbf{Allocation (reuse)} & None & None \\
            \textbf{Release} & Random(0x0-0xf)* & Increment \\
            \bottomrule
        \end{tabular}
        \begin{tablenotes}
            \item[*] \cc{OddEvenTags} support.
        \end{tablenotes}
    \end{threeparttable}
\end{table}

We evaluated the \sys-v2 gadget in the V8 JavaScript engine in two
environments: i) the standalone V8 JavaScript engine, and ii) the
Chromium application.
The V8 JavaScript engine runs as an independent process, reducing the
interference from the Android platform.
The Chromium application runs as an Android application, subject to
the Android's application management such as process scheduling and
thermal throttling.
The experiments were conducted with the V8 v12.1.10 and Chromium
v119.0.6022.0 release build.

We leveraged MTE random tagging schemes provided by the underlying
allocators~(\autoref{t:allocator}).
The standalone V8 used the Scudo allocator~\cite{scudo} (i.e., Android
default allocator), which supports 16 random tags for random tagging
and offers the \cc{OddEvenTags} option.
When \cc{OddEvenTags} is enabled, Scudo alternates odd and even
random tags for neighboring objects, preventing linear overflow
(i.e., \cc{OVERFLOW_TUNING}).
When \cc{OddEvenTags} is disabled, Scudo utilizes all 16 random tags
for every object to maximize tag entropy for use-after-free
detection (i.e., \cc{UAF_TUNING}).
By default, \cc{OddEvenTags} is enabled, while we evaluate both
settings.
Upon releasing an object, Scudo sets a new random tag that does not
collide with the previous one.
PartitionAlloc (i.e., Chrome default allocator) utilizes 15 random
tags and reserves the tag 0x0 for unallocated memory.
When releasing an object, PartitionAlloc increments the tag by one,
making the tag of the re-allocated memory address predictable.
However, in real-world exploits, it is challenging to precisely
control the number of releases for a specific address, thus we assume
the attacker still needs to leak the tag after each allocation.

For the evaluation, we constructed the \sys-v2 gadget in
JavaScript~(\autoref{fig:g2}) and developed MTE bypass
attacks as described in~\autoref{s:chrome-exploit}.
These attacks exploit artificial vulnerabilities designed to mimic
real-world renderer vulnerabilities, specifically linear
overflow~\cite{cve-2023-5217} and use-after-free~\cite{cve-2020-6449}.
We developed custom JavaScript APIs to allocate, free, locate, and
access the renderer object to manipulate the memory layout and trigger
the vulnerabilities.
It's worth noting that our evaluation shows the best-case performance
of MTE bypass attacks since real-world renderer exploits involve
additional overheads in triggering the vulnerabilities and controlling
the memory layout.

\begin{table}
    \centering
    \scriptsize
    \caption{Results of MTE bypass exploits against the V8 JavaScript engine}
    \label{t:v8-exploit}
    \begin{threeparttable}
        \begin{tabular}{llrr}
            \toprule
            \textbf{Vuln.} & \textbf{OddEvenTags} & \textbf{Accuracy} & \textbf{Time (s)} \\
            \midrule
            \textbf{Tag Leakage} & 0 (\cc{UAF_TUNING}) & 100/100 (100.00\%)  & 3.04 \\
            \textbf{Linear Overflow} & 0 (\cc{UAF_TUNING}) & 98/100 (98.00\%) & 13.52 \\
            \textbf{Linear Overflow} & 1 (\cc{OVERFLOW_TUNING}) & N/A & N/A \\
            \textbf{Use-After-Free}  & 0 (\cc{UAF_TUNING}) & 97/100 (97.00\%) & 10.66 \\
            \textbf{Use-After-Free}  & 1 (\cc{OVERFLOW_TUNING}) & 98/100 (98.00\%) & 6.09 \\
            \bottomrule
        \end{tabular}
    \end{threeparttable}
\end{table}

\begin{table}
    \centering
    \scriptsize
    \caption{Results of MTE bypass exploits against the Chromium application}
    \label{t:chromium-exploit}
    \begin{threeparttable}
        \begin{tabular}{lrr}
            \toprule
            \textbf{Vuln.} & \textbf{Accuracy} & \textbf{Time (s)} \\
            \midrule
            \textbf{Tag Leakage}     & 95/100 (95\%)   &  2.54 \\
            \textbf{Linear Overflow} & 97/100 (97\%)  &  16.11\\ 
            \textbf{Use-After-Free}  & 95/100 (95\%)  &  21.90 \\
            \bottomrule
        \end{tabular}
    \end{threeparttable}
\end{table}

\PP{V8 JavaScript Engine}
In the standalone V8 JavaScript engine, we evaluated the tag leakage
of the \sys-v2 gadget with \emph{cache eviction} and a
\emph{memory-based timer}.
For cache eviction, we used an L1 index-based random eviction set,
500 elements for \cc{slow[0]} and \cc{probe[PROBE_OFFSET]}, 300
elements for \cc{victim.length}.
The eviction performance of the random eviction set varies on each
run, so we repeated the same test 5 times and listed the best result.
The random eviction can be optimized with eviction set
algorithms~\cite{evset}.
We used a memory counter-based timer with a custom worker thread
incrementing a counter, which is equivalent to the
\cc{SharedArrayBuffer} timer~\cite{fantastic}. 
For all possible tag guesses (i.e., 0x0-0xf), we measured the access
latency of \cc{probe[PROBE_OFFSET]} after the gadget 256 times and
determined the guessed tag with the minimum average access latency as
the correct tag.

\autoref{t:v8-exploit} summarizes the MTE bypass exploit results in
V8.
For a single tag leakage, the gadget was successful in all 100 runs
(100\%), with an average elapsed time of 3.04 seconds.
MTE bypass exploits were evaluated over 100 runs for each
vulnerability and \cc{OddEvenTags} configuration (i.e., disabled (0)
and enabled (1)).
We excluded linear overflow exploit with \cc{OddEvenTags} enabled,
since the memory corruption is always detected with spatially adjacent
objects tagged with different tags and the attack would always fail.
The results demonstrate that the attacks were successful in over 97\%
of the runs, with an average elapsed time of 6 to 13 seconds.
In use-after-free exploits, enabling \cc{OddEvenTags} decreased the
average elapsed time by around 40\%, due to the decrease in tag
entropy from 16 to 8, doubling the chance of tag collision between
the temporally adjacent objects.

\PP{Chromium Application}
\label{s:eval:chromium}
In the Chromium application setting, we evaluated the \sys-v2 gadget
with \emph{cache flushing} and a \emph{\cc{SharedArrayBuffer}-based
timer}.
Unlike V8, random eviction did not effectively evict cache lines, so
we manually flushed the cache lines with \cc{dc civac} instruction.
We attribute this to the aggressive resource management of Android,
which can be addressed in the future with cache eviction algorithms
tailored for mobile applications.
To measure the cache eviction set overhead, we included the cache
eviction set traversals in all experiments, using the same cache
eviction configuration of the V8 experiments.
We measured access latency with a \cc{SharedArrayBuffer}-based timer
as suggested by web browser speculative execution
studies~\cite{leaky-page, spookjs}.
The MTE bypass exploits experiments were conducted in the same manner
as the V8 experiments.

\autoref{t:chromium-exploit} shows the MTE bypass exploit results in
the Chromium application.
The tag leakage of the \sys-v2 gadget in the Chromium application was
successful in 95\% of 100 runs, with an average elapsed time of 2.54
seconds.
With the MTE bypass exploits, success rates were over 95\% for both
vulnerability types, with an average elapsed time of 16.11 and 21.90
seconds for linear overflow and use-after-free, respectively.

\subsection{Linux Kernel Tag Leakage}
\label{s:eval:linux}

\begin{figure}[t]
    \centering
    \begin{subfigure}{0.48\columnwidth}
      \centering
      \includegraphics[width=\columnwidth]{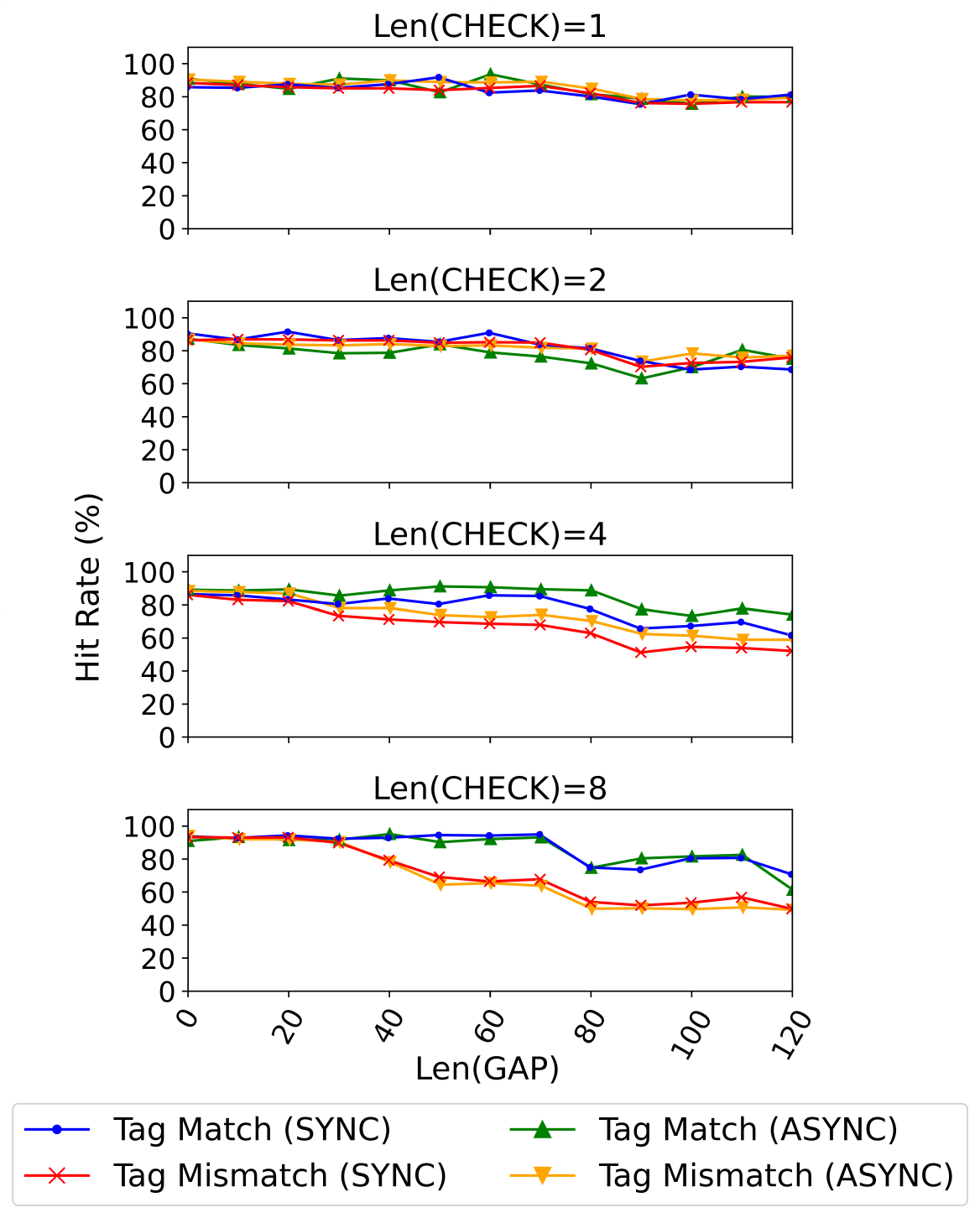}
      \caption{\sys-v1 (user space)}
      \label{fig:g1-kernel-eval}
    \end{subfigure}
    \hspace{0.3em}
    \begin{subfigure}{0.48\columnwidth}
        \centering
        \includegraphics[width=\columnwidth]{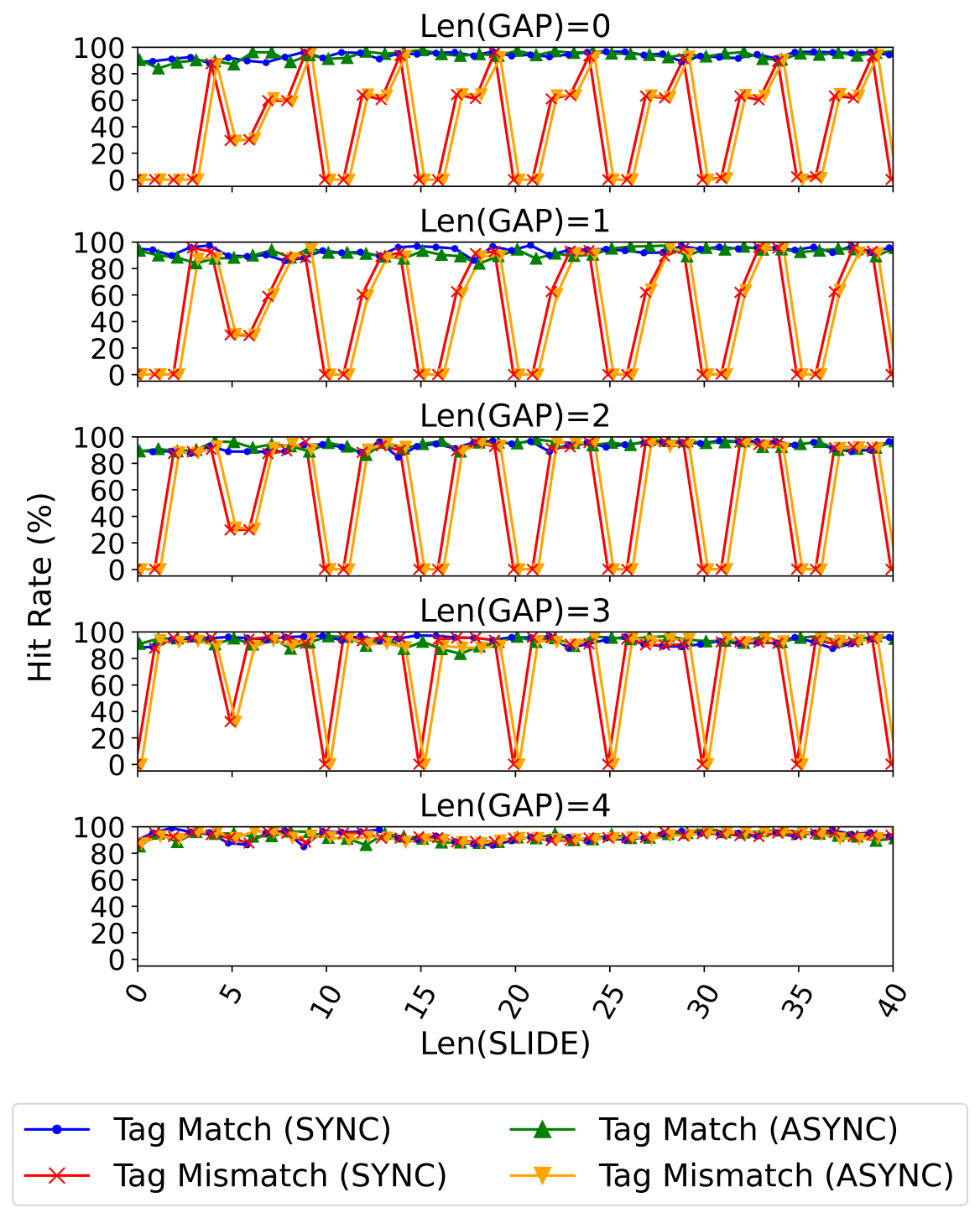}
        \caption{\sys-v2 (kernel space)}
        \label{fig:g2-kernel-eval}
    \end{subfigure}
    \caption{Kernel context evaluation of \sys gadgets}
    \label{fig:kernel-eval}
\end{figure}

The experiments were conducted on the Android 14 kernel v5.15 using
the default configuration.
We used 15 random tags (i.e., 0x0--0xe) for kernel objects, as tag 0xf
is commonly reserved for the access-all tag in the Linux
kernel~\cite{mte-kernel}.
The cache line eviction of kernel address \cc{cond_ptr} to trigger
the speculative execution was achieved by cache line
bouncing~\cite{cache-line-bouncing} from the user space.
For cache measurement, we utilized the virtual counter (i.e.,
\cc{CNTVCT_EL0}) to determine the cache hit or miss with the
threshold \cc{1.0}, which is accessible from the user space.
As the virtual counter has a lower resolution~(24.5MHz) than the CPU
cycle frequency~(2.4-2.9 GHz), the accuracy of the cache hit rate is
lower than the physical CPU counter-based measurements
in~\autoref{s:gadget}.
The access time was measured in the user space or kernel space,
depending on the experiment.

\PP{Kernel Context Evaluation}
We first evaluated whether \sys gadgets can leak MTE tags in the Linux
kernel context~(\autoref{fig:kernel-eval}).
We created custom system calls containing \sys-v1~(\autoref{fig:g1})
and \sys-v2~(\autoref{fig:g2}) gadgets and executed them by calling 
the system calls from the user space.
In \cc{CHECK}, we accessed the \cc{guess_ptr} that holds either the
correct or wrong tag \cc{Tg}.
In \cc{TEST}, \cc{test_ptr} pointed to either a kernel address or a
user space address, depending on whether the cache state difference
was measured in the kernel or user space.
When we leveraged a user space address as \cc{test_ptr}, we passed a
user buffer pointer to the kernel space as a system call argument and
accessed the pointer in \cc{TEST} using \cc{copy_to_user()}.
The user space address was flushed in the user space before the system
call invocation, and the cache state was measured after the system
call returned.
When we used a kernel address as \cc{test_ptr}, the cache flush and
measurement were performed in the kernel.
Each experiment measured the access time over 1000 runs.

When executing \sys-v1 in the kernel context, the MTE tag leakage was
feasible in both the kernel and user space, where the user space
measurement results are shown in~\autoref{fig:g1-kernel-eval}.
Compared to the user space gadget
evaluation~(\autoref{fig:g1-graph}), the kernel context required more
loads in \cc{CHECK} to distinguish the cache state difference.
Specifically, the cache state difference was discernible from 4 loads
in the kernel context, while the user space context required only 2
loads.
This can be attributed to the noises from the kernel to the user
space context switch overhead, such that the cache hit rates of the
tag match cases were lower (i.e., under 90\%) than the user space
gadget evaluation (i.e., 100\%).

When executing the \sys-v2 gadget in the kernel space, MTE tag leakage
was observed only in the kernel space~(\autoref{fig:g2-kernel-eval}).
When we measured the access latency of \cc{test_ptr} in the user
space, the gadget did not exhibit a cache state difference.
Although the \sys-v2 gadget might not be directly exploitable in the
user space, cache state amplification techniques~\cite{leaky-page,
hacky-racers} could be utilized to make it observable from the user
space.

\begin{table}
    \centering
    \scriptsize
    \caption{Results of MTE bypass exploits against the Linux kernel}
    \label{t:kernel-exploit}
    \begin{threeparttable}
        \begin{tabular}{lrr}
            \toprule
            \textbf{Vuln.} & \textbf{Accuracy} & \textbf{Time (s)} \\
            \midrule
            \textbf{Tag Leakage (\cc{artificial})} & 100/100 (100\%) & 0.12 \\
            \textbf{Tag Leakage (\cc{snd\_timer\_user\_read()})} & 100/100 (100\%) & 3.38 \\
            \midrule
            \textbf{Buffer Overflow (\cc{artificial})} & 100/100 (100\%) & 0.18 \\
            \textbf{Use-After-Free (\cc{snd\_timer\_user\_read()})} & 97/100 (97\%) & 6.86 \\
            \bottomrule
        \end{tabular}
    \end{threeparttable}
\end{table}

\PP{Kernel MTE Bypass Exploit}
We evaluated MTE bypass exploits in the Linux kernel with two \sys-v1
gadgets: an artificial \sys-v1 gadget with 8 loads in \cc{CHECK}
(i.e., \cc{artificial}) and a real-world \sys-v1 gadget in
\cc{snd_timer_user_read()}~(\autoref{code:kernel-gadget}).
The artificial gadget evaluates the best-case performance of MTE
bypass attacks, while the \cc{snd_timer_user_read()} gadget
demonstrates real-world exploit performance.
Both gadgets were triggered by invoking the system call containing
the gadget from the user space, leveraging a user space address as
\cc{test_ptr}, and measuring the access latency of \cc{test_ptr} in
user space.

We conducted a tag leakage attack and MTE bypass attack for each
gadget.
For the MTE bypass attack, we synthesized a buffer overflow
vulnerability.
Each gadget dereferenced the vulnerable pointer (i.e., \cc{guess_ptr})
to trigger tag checks; an out-of-bounds pointer and a dangling pointer
for the buffer overflow and use-after-free exploits, respectively.
The exploit methodology followed the process described
in~\autoref{s:appendix:linux}.

\autoref{t:kernel-exploit} summarizes the MTE bypass exploit results.
For a single tag leakage, the gadgets successfully leaked the correct
tag in all 100 runs (100\%), with an average elapsed time of 0.12
seconds in the artificial gadget, and 3.38 seconds in the
\cc{snd_timer_user_read()} gadget.
The MTE bypass exploit for the artificial \sys-v1 gadget was
successful in all 100 runs (100\%), with an average elapsed time of
0.18 seconds.
Regarding the MTE bypass exploit for the \cc{snd_timer_user_read()}
gadget, the success rate was 97\% with an average elapsed time of 6.86
seconds.
As the \cc{snd_timer_user_read()} gadget involves complex kernel
function calls and memory accesses, the performance of the MTE bypass
exploit is slightly lower compared to the artificial gadget.
Nevertheless, it still demonstrates a high success rate within a
reasonable time frame.

\section{Related work}
\label{s:relwk}

\PP{MTE Security Analysis}
Partap et al.~\cite{mte-analysis} analyzed the software-level MTE
support in real-world memory allocators.
Google Project Zero~\cite{pz-mte} explored speculative execution
attacks against MTE hardware for the first time.
\cc{StickyTags}~\cite{sticky-tags} identified an MTE tag leakage
gadget (which is similar to \sys-v1) and proposed a deterministic
tagging-based defense that does not utilize random tags due to the
potential tag leakage.
Compared to \cc{StickyTags}, our work identified a new type of MTE tag
leakage gadget, \sys-v2, and analyzed the root cause of both \sys-v1
and \sys-v2 gadgets.
We also demonstrated the real-world exploitation of \sys gadgets in
Google Chrome and the Linux kernel and proposed new defense mechanisms
to mitigate the security risks posed by \sys gadgets.
While \cc{StickyTags} proposed deterministic tagging due to the
potential
tag leakage, our work focuses on hardening the random tagging-based
MTE defense, which are developed by major vendors including
Google~\cite{pixel}, the Linux kernel~\cite{hw-kasan}, and secure
operating systems~\cite{trustonic,optee,grapheneos}.

\PP{Speculative Attacks on Protection Mechanisms}
Speculative probing~\cite{blindside} suggested that speculative
execution can be used to probe address mappings and bypass address
space layout randomization (ASLR).
PACMAN~\cite{pacman} identified speculative gadgets that leak Pointer
Authentication Code (PAC).
ARMv8.6 \cc{FEAT_FPAC} mitigates PACMAN attacks by authentication and
memory access, allowing all memory accesses regardless of the
authentication result~\cite{pacman-mitigation}.
MTE tag leakage can also be mitigated by separating tag check and
memory access in the hardware, not allowing tag check results to
affect memory access.

\PP{Transient Execution Attacks}
Transient execution attacks exploit micro-architectural behaviors to
leak secret information.
Researchers have analyzed various micro-architectural implementations
including speculative
execution~\cite{spectre,meltdown,foreshadow,retbleed}, memory
disambiguation prediction~\cite{spectre-v4,spoiler,fallout}, and CPU
internal buffers~\cite{lvi,ridl}.
Recent attacks exploited data prefetching behaviors to leak secret
information or construct covert
channels~\cite{prefetcher-attack,augury,afterimage,fetchbench}.
Compared to these attacks, we identified for the first time that data
prefetching behaviors can also be exploited to leak hardware
exceptions, such as tag check faults~(\autoref{s:gadget1}).

\section{Conclusion}
\label{s:conclusion}

This paper explores the potential security risks posed by speculative
execution attacks against ARM Memory Tagging Extension (MTE).
We identify new MTE oracles, \sys-v1 and \sys v2, capable of leaking
MTE tags from arbitrary memory addresses.
\sys gadgets can bypass MTE-based defense in real-world systems,
including Google Chrome and the Linux kernel.
Our findings provide significant insights into the design and
deployment of both memory tagging-based hardware and software
defenses.

\begin{footnotesize}
    \bibliographystyle{abbrvnat}
    \bibliography{p,sslab,conf}        
\end{footnotesize}

\newpage
\appendices
\begin{figure}[t]
  \centering
  \includegraphics[width=0.5\columnwidth]{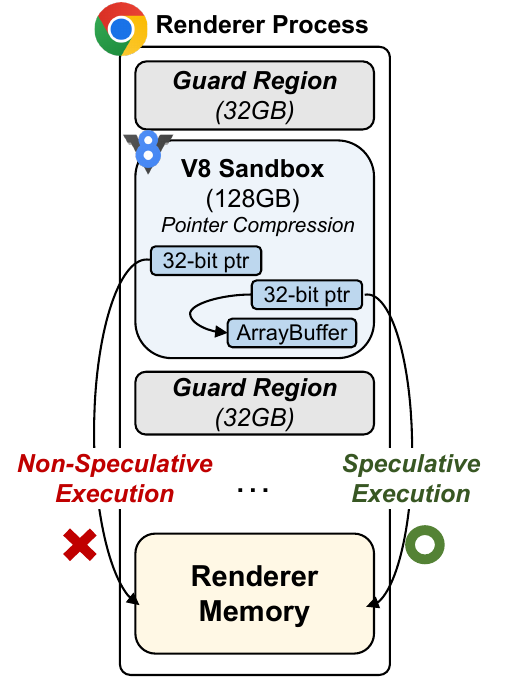}
  \caption{Speculative V8 sandbox escape}
  \label{fig:v8-sandbox}
\end{figure}
\begin{figure}[t]
  \centering
  \input{code/g2-js.asm}
  \caption{Turbofan optimized assembly of \sys-v2 gadget}
  \label{code:js-gadget-asm}
\end{figure}

\begin{figure}[t]
  \centering
  \begin{subfigure}{0.8\columnwidth}
    \centering
    \include{code/tiktagleak.js}
    \vspace{-1.5em}
    \caption{Tag Leakage}
    \label{code:js-tiktagleak}
  \end{subfigure}
  \begin{subfigure}{0.49\columnwidth}
    \centering
    \include{code/oob.js}
    \vspace{-1em}
    \caption{Spatial bugs}
    \label{code:chrome-oob}
  \end{subfigure}
  \begin{subfigure}{0.49\columnwidth}
    \centering
    \include{code/uaf.js}
    \vspace{-1em}
    \caption{Temporal bugs}
    \label{code:chrome-uaf}
  \end{subfigure}

  \caption{MTE bypass attacks against Chrome}
  \label{fig:chrome-sys-exploit}
\end{figure}

\begin{figure}[t]
  \begin{subfigure}{0.8\columnwidth}
    \centering
    \include{code/kernel-oob.c}
    \vspace{-2em}
    \caption{Spatial bugs}
    \label{code:kernel-oob}
  \end{subfigure}
  \\
  \begin{subfigure}{0.8\columnwidth}
    \centering
    \include{code/kernel-uaf.c}
    \vspace{-2em}
    \caption{Temporal Bugs}
    \label{code:kernel-uaf}
  \end{subfigure}
  \caption{Exploiting memory corruption bugs in the Linux kernel. Branch training and cache control are omitted for simplicity.}
  \label{fig:kernel-mte-exploit}
\end{figure}

\begin{figure*}[t]
  \centering
  \begin{subfigure}{\textwidth}
    \centering
    \includegraphics[width=\textwidth]{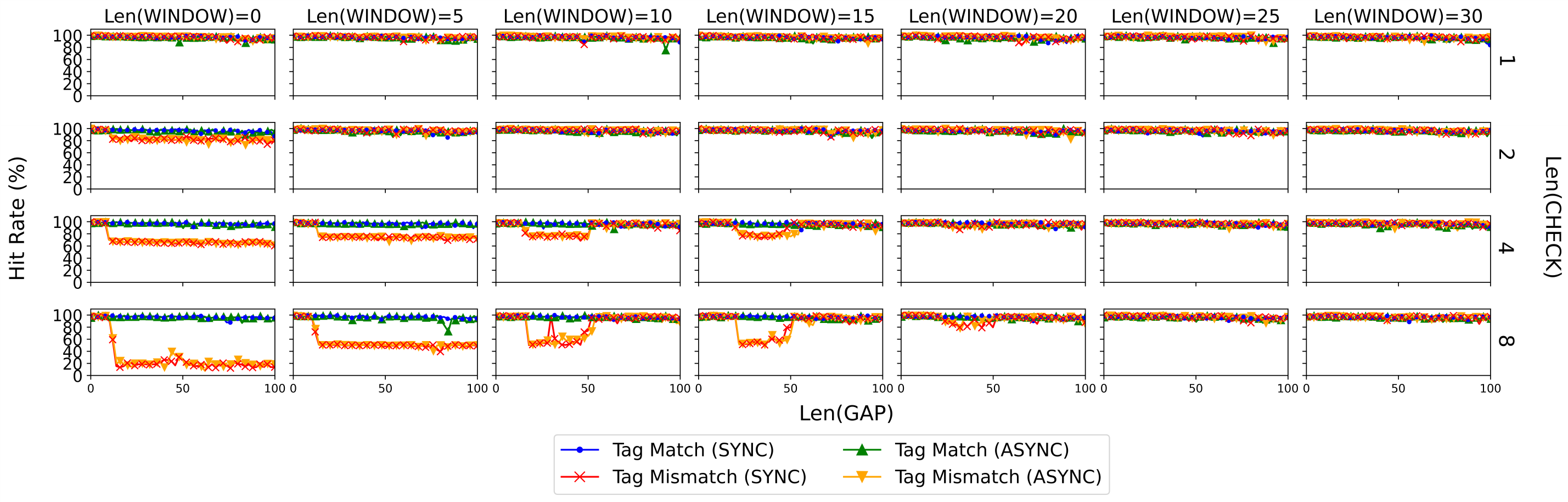}
    \caption{\cc{TEST=INDEP_LD}}
    \label{fig:g1-indep-ldr}
  \end{subfigure}
  \\
  \begin{subfigure}{\textwidth}
    \centering
    \includegraphics[width=\textwidth]{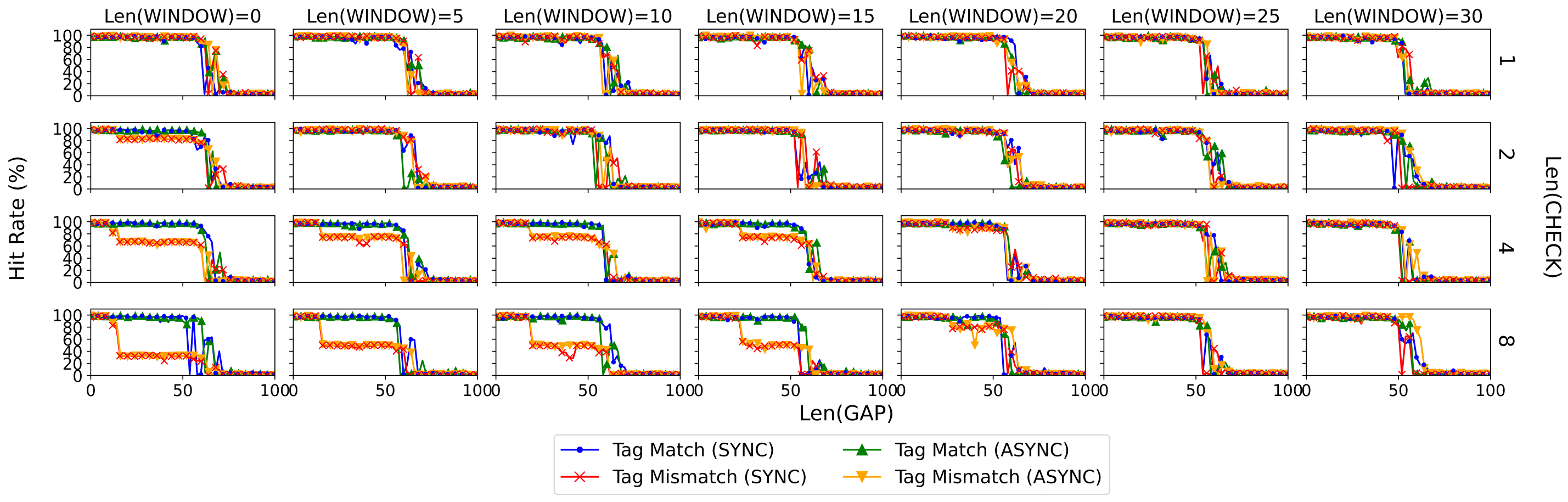}
    \caption{\cc{TEST=INDEP_ST}}
    \label{fig:g1-indep-str}
  \end{subfigure}
  \caption{\sys-v1 experiment by varying the \cc{BR} and \cc{CHECK}
  instruction distance (i.e., \cc{WINDOW})}
  \label{fig:g1-gap}
\end{figure*}

\section{Speculative V8 Sandbox Escape}
\label{s:appendix:v8-sandbox}
The V8 sandbox~\cite{v8-sandbox} isolates JavaScript from the rest of
the renderer by \emph{pointer compression} and \emph{guard
regions}~(\autoref{fig:v8-sandbox}).
Pointer compression enforces that all JavaScript pointers are
compressed into a 32-bit or 35-bit index to represent a maximum 128
GB memory in Android, where memory addresses are calculated by adding
the index to the sandbox base address.
Guard regions are placed at the start and end of the V8 sandbox with
32 GB size each, preventing potential out-of-bound access (i.e.,
worst case 32-bit index for the 8-byte element array).
Thus, any native memory access instructions executing JavaScript are
strictly limited to accessing the 4~GB boundary, and thus it cannot
access the renderer's memory.

Despite this strong memory isolation, we found the vulnerable case
that the V8 sandbox does not guarantee complete isolation in the
speculative paths.
This vulnerable case is in the bound check implementation of
\cc{TypedArray}.
\cc{TypedArray} in JavaScript is an array type, supporting up to 32~GB
array size~\cite{v8-max-size}.
The length of \cc{TypedArray} is represented as 35 bits in V8, stored
as a 64-bit field.
\cc{TypedArray} can be accessed using the index, which is a 64-bit
value.
Every \cc{TypedArray} access using the index is preceded by a bound
check, which compares the index against the length of the array.
Specifically, given a 64-bit index, the bound check compares the whole
64-bit index against the 35-bit length.
If the index is smaller than the length, the access is allowed;
Otherwise, the access is denied.

The problem occurs when the bound check for the \cc{TypedArray} access
is speculatively executed.
In this case, even if the index is larger than the array length, the
access is still speculatively executed.
This allows the untrusted JavaScript code to access beyond its
restricted region, thereby escaping V8's sandbox.
To the best of our knowledge, this speculative sandbox escape
vulnerability was first discovered by this work, and we accordingly
reported this to the V8 security team in December 2023.

\section{V8 Turbofan Optimized Assembly}
\label{s:appendix:v8asm}

\autoref{code:js-gadget-asm} shows the Turbofan optimized assembly of
\sys-v2 gadget in V8 JavaScript engine~(\autoref{code:js-gadget}).
\cc{BR} block contains the \cc{slow} branch at line 9-10. 
\cc{CHECK} block contains the \cc{PROBE_OFFSET} store and load
instructions at line 31-34, which are 2 machine instructions
distance, satisfying the requirement of \sys-v2 gadget (i.e., to be
within the 5 instruction dispatch window).
\cc{TEST} block contains the load instruction using the forwarded
value (i.e., \cc{val}) at line 39.

\section{Chrome V8 MTE Bypass Attack}
\label{s:appendix:v8-exploit}

\autoref{fig:chrome-sys-exploit} shows examples of \sys gadget-based
MTE bypass attacks in the Chrome.
A JavaScript function \cc{TikTagLeak()}~(\autoref{code:js-tiktagleak})
leaks the tag of the renderer memory with the \sys-v2 gadget.
This function brute-forces \cc{TikTag_v2()} to check \cc{Tg} against
the tag \cc{Tm} of \cc{target_addr}.
The assumption here is that the attacker implements \cc{ObjToIdx()},
which returns \cc{idx}, such that \cc{\&victim[idx]} points to the
\cc{target_addr}.
With the leaked tags, the attacker can exploit spatial bugs
(\autoref{code:chrome-oob}) and temporal bugs
(\autoref{code:chrome-uaf}) in the renderer without raising a tag
check fault.
The spatial bugs example is based on a linear buffer overflow
vulnerability in the renderer process,
CVE-2023-5217~\cite{cve-2023-5217}, while the temporal bugs example
is based on a use-after-free vulnerability in the renderer process,
CVE-2020-6449~\cite{cve-2020-6449}.
Both examples follow the same pattern: the attacker allocates
renderer objects, leaks their tags, and triggers memory corruption
only when the tags match.

\section{Linux Kernel MTE Bypass Attack}
\label{s:appendix:linux}

\autoref{fig:kernel-mte-exploit} demonstrates MTE bypass attacks
using the \sys gadgets in the Linux kernel.
\autoref{code:kernel-oob} illustrates a spatial bug exploit,
leveraging a kernel buffer overflow vulnerability~(e.g.,
CVE-2022-0185~\cite{cve-2022-0185}), and \autoref{code:kernel-uaf}
illustrates a temporal bug exploit, leveraging a kernel
use-after-free vulnerability (e.g.,
CVE-2019-2215~\cite{cve-2019-2215}).
Functions prefixed with \cc{Sys} denote system calls responsible for
kernel object allocation, deallocation, or triggering the \sys
gadgets or memory corruption.
Each vulnerable system call (i.e., \cc{SysExploitOOB()} and
\cc{SysExploitUAF()}) has their counterpart in \sys gadget (i.e.,
\cc{SysTikTagOOB()} and \cc{SysTikTagUAF()}) that leaks 
the tag check result through user space buffer (i.e., \cc{ubuf}).

\section{\sys-v1 additional experimental results}
\label{s:appendix:g1-gap}
\autoref{fig:g1-gap} shows the \sys-v1 gadget experiment results by
varying the number instructions (i.e., \cc{orr}) between the \cc{BR}
and \cc{CHECK} instructions (i.e., \cc{WINDOW}).
We experimented with the \sys-v1 gadget in the same setting as
in~\autoref{s:gadget1}, but with the length of \cc{WINDOW} varying
from 0 to 30.
We tested under two conditions: \cc{TEST} containing a load
instruction~(\autoref{fig:g1-indep-ldr}) or a store
instruction~(\autoref{fig:g1-indep-str}). 
In both cases, the \sys-v1 gadget showed cache hit rate difference
when \cc{WINDOW} is less than 25.
The results were consistent when the \cc{orr} instructions in
\cc{WINDOW} had no dependency, or had dependencies to \cc{CHECK}, or
even when they were replaced with \cc{nop} instructions.
Therefore, we think the time window between \cc{BR} and \cc{CHECK}
requires around 25 instructions to trigger the tag check fault.
As the affected core~(i.e., Cortex-X3) has a 6-instruction dispatch
window~\cite{x3-decode}, we think the time for the \sys-v1 to fetch
the \cc{WINDOW} instructions is around 5 cycles.
We think that if \cc{CHECK} triggers tag check faults within 5 cycles
from \cc{BR}, the CPU reduces the speculation execution and data
prefetching, as described in~\autoref{s:gadget1}.

\end{document}